\documentclass[times,onecolumn,review,nopreprintline]{elsarticle}
\usepackage{lipsum}
\usepackage{jasr}
\usepackage{framed, multirow}
\usepackage{footnote}
\usepackage{tablefootnote}
\DeclareUnicodeCharacter{2212}{\textminus}
\usepackage{gensymb}
\usepackage{graphicx}
\usepackage{parskip}
\usepackage{amsmath}
\usepackage{siunitx}
\usepackage{multirow}
\usepackage{wrapfig}
\usepackage{enumitem}
\usepackage{subcaption}

\usepackage{float}
\DeclareUnicodeCharacter{2009}{\,} % Thin space

\usepackage[x11names]{xcolor}
\definecolor{new color}{rgb}{.8,.349,.1}
\usepackage[citebordercolor=white,citecolor=blue,colorlinks=true,linkcolor=blue]{hyperref}
\journal{Advances in Space Research}
\usepackage{natbib}
\usepackage{url}
\biboptions{authoryear}

\begin{document}

\verso{ C. Muthumariappan \textit{etal}}

\begin{frontmatter}

\title{Infrared properties of Planetary Nebulae with PG1159 central stars\tnoteref{tnote1}}%
%\tnotetext[tnote1]{This is an example for title footnote coding.} 

\author[1]{C. \snm{Muthumariappan}\corref{cor1}}
\cortext[cor1]{Corresponding author: 
  Tel.: +91 80 25530672; 
  fax: +91 80 25534043;}
\author[1,2]{K. \snm{Khushbu}}
\author[3]{V. \snm{Kerni}}\fnref{fn1}
%\fntext[fn1]{This is author footnote for second author.}
%% Third author's email
%\ead{muthu@iiap.res.in}
%\author[2]{Given-name4 \snm{Surname4}}

\address[1]{Indian Institute of Astrophysics, Koramangala, Bangalore, 560 034, India} 
\address[2]{Pondicherry University, R.V. Nagar, Kalapet, 605014, Puducherry, India}
\address[3]{Indian Institute of Technology, Roorkee, 247 667, India \\}
%\address[3]{Vainu Bappu Observatory, Indian Institute of Astrophysics, Kavalur, Alangayam, 635 701, India}
\received{ }
\finalform{ }
\accepted{}
\availableonline{}
\communicated{}

\begin{abstract}
%%%
We study the properties of 26 PNe with PG1159-type central stars known till date and compare them with the  properties of PNe having
[WR], $wels$ and hydrogen-rich central stars published earlier. We use archival photometric measurements of $2MASS$ for near-IR analysis 
and $WISE$ and $IRAS$ data for mid- and far-IR analysis and derive the IR properties of PG1159-PNe. We analyze the IR colour-colour diagrams of 
PG1159-PNe and compare them with the other three groups of PNe. Similar to the [WR]-PNe, many PG1159-PNe also show large amount of near-IR emission 
from the hot-dust component but their AGB dust is relatively cooler. We also report here the dust colour temperatures, dust masses, dust-to-gas 
mass ratios, IR luminosities and IR excess of PG1159-PNe and plot them against their surface H$\beta$ brightness (age) and compare them with the 
distribution of other groups of PNe. The IR luminosity and dust temperature show strong correlation with surface H$\beta$ brightness, however, 
the dust-to-gas mass ratio and IR 
excess do not show any trend. While the mean dust mass has a lower value for PG1159-PNe, in compared to other groups, the average dust-to-gas mass 
ratio is found to be marginally larger for PG1159-PNe. An analysis of the number distribution of different groups of PNe against surface H$\beta$ 
brightness shows that a) younger [WR]-, $wels$- and normal-PNe have a similar distribution indicating that they all have evolved from the AGB in a 
similar way, b) while there is an overlap of surface H$\beta$ brightness between [WR]- and PG1159-PNe, showing an evolutionary connection between 
them, there exists a significant gap between the values derived for $wels$- and PG1159-PNe.

%%%%
\end{abstract}

\end{frontmatter}

%% For linenumbers
% \linenumbers

\section{Introduction}
\label{sec1}
%Please use \verb+elsarticle.cls+ for typesetting your paper. Additionally,
%make sure not to remove the package \verb+jasr.sty+ already included in the
%preamble:
%\begin{verbatim} 
%  \usepackage{jasr}
%\end{verbatim}
The planetary nebula (PN; plural PNe) phase occurs when the low- and intermediate-mass stars with initial masses 0.8 to 8 $M_{sun}$ evolve from the AGB phase before 
attaining white-dwarf configuration as the end product. Depending on the hydrogen content in their atmospheres, the central stars of PNe (CSPNe) was classified into 
hydrogen-rich or hydrogen-poor (\cite{1991IAUS..145..375M}). While most CSPNe are  hydrogen-rich (H-rich, normal-PNe hereafter; \cite{2006IAUS..234..127T}, about 30$\%$ 
of the CSPNe belong to the hydrogen-poor (H-poor) category, where hydrogen is depleted in their stellar atmospheres and they have helium and carbon as the most abundant 
elements. A sizeable number of CSPNe (about 7$\%$) are known to exhibit Wolf-Rayet type spectrum \citep{2008ASPC..391...83C} (hereafter [WR] stars and the PNe with [WR] 
CSPNe are [WR]-PNe) which show H-poor atmospheres with strong and broad emission lines of highly ionized carbon, oxygen and helium. [WR] stars are also subdivided into the early type [WRE] and late type [WRL] \citep{2003A&A...403..659A}. There is also a group of CSPNe which 
show H-poor atmosphere and weak and narrow emission lines at the same wavelengths as those shown by [WR] stars, which are called the 'weak emission lines stars' ($wels$; 
PNe with $wels$ CSPNe are $wels$-PNe hereafter). PG1159 stars form another group of H-poor post-AGB stars with similar surface abundances to other H-poor CSPNe. While 
[WR] stars and $wels$ stars are always surrounded by PNe; only a fraction of PG1159 stars have PNe around them,  which are usually faint (PG1159-PNe hereafter). \\
PG1159 stars represent post-AGB evolution with the hottest known H-poor atmospheres (T$_{eff}$ up to 200,000 K)\citep{Werner_2006, 10.1093/mnras/stz1994}. They are named after the proto-type star PG1159-035, a 
faint blue object of magnitude $14.5$ by \cite{pg1159winget} in the Palomar green survey \citep{greensurvey}. In the H-R diagram, PG1159 stars occupy a region with low 
luminosity and high surface temperature\citep{Werner_2006, 2008ASPC..391..209D}. Their optical spectrum is dominated by He II and C IV absorption lines \citep{Werner1993}. It is believed that PG1159 stars have experienced 
re-ignition of helium, resulting in helium flash (thermal pulse \cite{thermalpulse}), which dredged up the stellar atmospheres with the inter-shell abundance of carbon 
and oxygen metals, like other H-poor stars \citep{2000A&A...362.1008G}. Hence, H-poor CSPNe are the test beds of helium-burning CSPNe models \citep{2001A&A...377.1007G}. 
Some of these 
PG1159 stars might still be fusing helium. There is also a sub-group of PG1159 CSPNe called hybrid PG1159 \citep{ref10} which show hydrogen in absorption. With the high surface temperature and gravity, PG1159 remained a unique stellar object until it was discovered that the central 
star of the PN K 1-16 is a pulsating variable star similar to PG1159-035. Subsequently, more objects which are similar to PG1159-035 and K 1-16 were discovered with an 
extended Palmoar survey, resulting in a new class of extremely hot pulsators. Other characteristics of PG1159 stars are that they are non-radial g-mode pulsators with 
varying pulsation periods and are given a sub-class of GW Variable stars (or GW Vir), for example, SDSS J0349−0059 \citep{2012MNRAS.426.2137W}. In his work, \cite{Werner1993}
has introduced multiple sub-classes in the PG1159 class of stars based on these variable pulsations. \\

The evolutionary path of low- and intermediate-mass stars which give birth to H-poor CSPNe should have some difference with the path followed by H-rich CSPNe. How and 
when their evolutionary paths differ is not yet well understood. Importantly, the evolutionary connections between different types of H-poor CSPNe are not fully known. PNe around H-rich and H-poor CSPNe should have this information imprinted with them; hence their 
properties must be probed in detail. The chemical composition and other properties of 30 [WR]-PNe and 18 $wels$-PNe were studied and compared to a sample of normal-PNe 
by \cite{girard:hal-00119849} . \cite{2001A&A...377.1007G} analyzed the IR properties of 49 [WR]-PNe and addressed their dust content using near- and mid-IR photometric and spectroscopic observations gathered from the literature. A recent study on the physical properties of [WR]-PNe with its sub-types can also be seen in \cite{2023RAA....23i5021A} . A uniform IR analysis of known PNe surrounding 99 [WR]- and 67 $wels$- central stars was made by \cite{cmuthu} (MP20, hereafter), and their properties were compared with a randomly selected sample of 100 PNe with H-rich central stars. Nebular and dust properties 
derived by them using $IRAS$, $WISE$ and $Akari$ data showed several differences from which they addressed the evolutionary connections between different groups of PNe. 
They argued that [WR]-PNe and $wels$-PNe are two different groups of PNe, and they do not form an evolutionary sequence, though they share many similar properties. Their study also found very old PNe only in the normal-PNe group and [WR]- and $wels$-PNe are not seen as very old PNe, suggesting that they evolve to  group of PNe
with different type of H-poor central stars, possibly PG1159. \\

Though it is believed that [WR] stars can turn into PG1159, possible evolutionary link(s) between PG-1159 stars and [WR] and $wels$ stars are still unknown. 
 It is essential to study the infrared and other related properties of all known PG1159-PNe and compare them with the [WR]- and $wels$-PNe properties in order to address the 
evolutionary links between different H-poor CSPNe. 
%The proposed evolutionary sequence of \textit{Post-AGB} $\rightarrow$ \textit{[WR]} (\textit{[WRL]} $\rightarrow$ 
%\textit{[WRE]}) stars $\rightarrow$ \textit{wels} stars $\rightarrow$\textit{PG/hybrid-1159} stars $\rightarrow$ \textit{DO %type white dwarfs}(WDs) is required to be checked. --> to discussion
 We aim to do this by investigating the IR properties of PG1159-PNe and their relation with the IR properties derived for [WR]- and $wels$-PNe by MP20 and also aim to make an age analysis for different groups of H-poor PNe. Nebulae around many PG1159 stars were dispersed into the ISM, and this study can be made with the limited number of PNe around these stars. We compile 26 known PNe with PG1159 central stars, and we use the IR colour-colour diagrams of PG1159-PNe derived using photometry at $2MASS$, $IRAS$ and $WISE$ bands and also derive their nebular and 
dust properties from $IRAS$ fluxes at 25- and 60$\mu$m bands. We also invoke a statistical analysis of H$\beta$ surface brightness distribution of all groups of PNe to constrain the origin of H-poor PNe and to understand the evolutionary connection of PG1159 stars with the [WR] and $wels$ stars in their journey towards white dwarf. Though the interest of the community is further enhanced by the presence of pulsations in PG1159 CSPNe to study their internal structure, it is beyond the scope of this work, and hence it is not addressed here.\\

We organize the paper in the following way: Section \ref{sec2} provides the data sample and computation of IR properties; Section \ref{sec3} presents the results we obtained from the i) colour-colour diagrams in the near and mid-IR bands for [WR]-,$wels$-, PG1159- and normal-PNe, ii) gas and dust properties derived for these groups of PNe, iii) statistical analysis of $S_{H\beta}$ for different groups of PNe with H-poor central stars. A discussion on the dust-to-gas mass ratios in PNe and on the origin of PNe with H-rich and H-poor central stars are given in Section \ref{sec4}. Our conclusions are listed in Section \ref{sec5}.
%Prototypical PG 1159 − 035 are grouped in the E sub-class (or emission lines). Other sub-classes introduced named A, 1gE, 1gEp, Ep and x \footnote{x represents A, E class}.

\section{Data Sample and computation of IR parameters}
\label{sec2}
We used the IR photometric data obtained from the archives for this study. The $Two$ $Micron$ $All$ $Sky$ $Survey$ ($2MASS$) \citep{2mass} data was used to compute the near-IR fluxes from the magnitude measurements available at different wavebands. Mid- and far-IR fluxes were computed from the stellar magnitudes obtained from the $Wide-field$ $Infrared$ $Survey$ $Explorer$  ($WISE$; \cite{wisepaper}, \cite{allwise}) at 3.4-, 4.2-, 12- and 22$\mu$m bands and the fluxes at 12-, 25-, 60- and 100$\mu$m bands of $InfraRed$ $Astronomical$ $Satellite$ ($IRAS$; \cite{iraspaper}). The IR data taken for this study are available at the NASA/IPAC Infrared Science Archive $^{a}$. $Akari$ data were not available for most of the sources, and we don't use them here as they form poor statistics. All the archival data we have used for this study have definite error measurements. They were used to estimate the errors associated with the derived parameters. \\

  We searched for the confirmed candidate of PG1159 stars with PNe around them in the literature. There are 67 objects which were classified as PG1159 stars; among them 26 sources (39$\%$) have PNe. A list of  22 PG1159-PNe and 4 hybrid PG1159-PNe are given in Table 1, which makes our sample (WeSb 3 is a doubtful PG1159-PN hence it is not included \citep{2011A&A...526A...6W} ). Most of them come from the catalogue of spectral classification of CSPNe given by \cite{2011A&A...526A...6W}. References for PN candidates which have hybrid-PG1159 central stars are given in the table. We have used the nebular $H{\beta}$ fluxes and electron densities of PNe, required for our investigation given in \cite{stasinskaed}.
Optical size of the nebula the values of $E(B-V)$ and the distances for all the PG1159-PNe in our sample are from \cite{frew2016halpha}. Though distances for some sources are also given at Gaia DR3 release, they are not available for all objects, and as discussed in MP20 Gaia distances match reasonably well 
with the values derived by \cite{frew2016halpha}. Some objects in our sample do not have IR data in some bands and/or they do not have the nebular $H{\beta}$ flux 
and/or density and/or distance measurements. However, the available data on the PG1159-PNe in our sample are maximally utilized: a study is performed for an 
object in the following sections if the required data for that particular study is available with measurement error.
The $2MASS$ photometric fluxes were corrected for interstellar extinction using the relations given by \cite{1985MNRAS.213...59W} and
the $E(B-V)$ values given in Table 1. The nebular and dust properties of PG1159- and hybrid PG1159-PNe were computed using the methodology discussed in 
detail by MP20 (section 4.3); which we brief here:  
  
 The characteristic temperature of thermal equilibrium dust in a PN can be derived using its far-IR continuum fluxes at $IRAS$ 25- and 60$\mu$m bands, 
 where the contribution from dust emission features and atomic line emission are minimal. These far-IR fluxes were fitted with a modified blackbody 
 function assuming an emissivity exponent of 1.0 \citep{Muthumariappan_2006}. PNe are optically thin at far-IR wavelengths which makes the determination of 
 $T_{d}$ easy. The dust temperature of a PN derived in this way is also called the dust colour temperature 
 ($T_{d}$; \cite{stasinskaed}). We derive the dust colour temperatures of our sample PNe for which the $IRAS$ 25- and 60$\mu$ fluxes are known with 
 the data quality flag 3. The  distance-independent parameter, the nebular H${\beta}$ surface brightness ($S_{H\beta}$; \cite{stasinskaed}, 
 \cite{2000A&A...362.1008G}, MP20),  was derived from the H$\beta$ flux and the optical size of the nebula. $S_{H\beta}$ is a good indicator of the age of a PN: as the nebula evolves, it expands and the $S_{H\beta}$ value decreases \citep{stasinskaed}.

 IR luminosity ($L_{IR}$) of a PN was estimated from its $IRAS$ fluxes at 25- and 60$\mu$m bands, which was fitted by a modified blackbody flux 
 distribution function. The curve was then integrated above 1$\mu$m to find the total $L_{IR}$ from continuum dust, taking the distance 
 into account. We estimated the values of $L_{IR}$ and their errors for PG1159- and hybrid PG1159-PNe (the errors mainly come from the distance measurements). The IR excess (IRE) and their errors were estimated from the $L_{IR}$ values and their reddening corrected nebular H$\beta$ fluxes 
 and using the relation (equation 6) given by \cite{stasinskaed}. IRE were calculated for PNe if they have both $IRAS$ far-IR and 
 H$\beta$ fluxes available with their uncertainties. 

 The dust mass $m_{d}$ and the dust-to-gas mass ratios $m_{d}/m_{g}$ were estimated by assuming an optically thin limit for PNe in 
 the mid-and far-IR region. For a given distance, the dust mass was derived using the continuum flux at $IRAS$ 60$\mu$ band and the $T_{d}$ 
 value. The values of $T_{d}$, $IRAS$ 60$\mu$ flux, H$\beta$ flux and the electron density taken from the literature were used to obtain the 
 dust-to-gas mass ratio. Errors associated with $m_{d}$ and $m_{d}/m_{g}$ were also estimated. \\
 
\begin{table*}
%\begin{minipage}{120mm}
\label{tbl1}
\caption{List of PG1159-PNe and Hybrid-PG1159-PNe. Origin of the sources: 1) \cite{2023A&A...676A...1W}, 2) \cite{2023MNRAS.521..668B}, 3) \cite{2004A&A...424..657W}.
Other sources are taken from \cite{2011A&A...526A...6W}. Diameter ($\theta$), distances and E(B-V) values are from \cite{frew2016halpha}. F$_{H\beta}$ and  $n_{e}$ are from \cite{stasinskaed}}
\vspace{1cm}
\begin{tabular}{lrrrrrrr}
\hline 
\hline 
PNG & other & $F_{H\beta}$ &  $\theta$ & $n_{e}$     & $E(B-V)$  & Distance & IR data $^{a}$    \\
No.   & name  &  $(10^{-12} erg/cm^2/s)$ & (arcsec) & ($cm^{-3}$)  &         &  (kpc)     & \small{(2MASS/WISE/IRAS)}     \\
\hline
            &          &      &      &              &              &                &            \\
PG1159-PNe  &          &      &      &              &              &                &             \\
%$028.0+10.2$  & $WeSb3 ^{3}$   &--    &--     &--           &--            &--              &--            \\
042.5-14.5  & NGC 6852 & $2.90\pm0.39$ &14 &-- & $0.14\pm0.07$ & $4.65\pm1.33$ &  yes/yes/yes   \\
046.8+03.8  & Sh 2-78  & $8.46\pm0.42$ &300  &-- &  $0.01\pm0.002$ &   $1.05\pm0.20$   & yes/no/no         \\
059.7-18.7  & A 72     &$2.20\pm$-- &14 &-- & $0.05\pm0.03$ &$2.56\pm0.73$ &  yes/yes/yes  \\
062.4+09.5  & NGC 6765 & $1.73\pm0.41$ &19 &-- &$0.19\pm0.17$ & $3.70\pm1.28$ &  yes/yes/yes  \\
$070.5+11.0$  & Kn 61 $^{2}$   &$0.12\pm$--    &96   &--          & $0.15\pm0.03$   &   $5.41\pm1.51$  & no/no/no  \\
080.3-10.4  & MWP1     &$6.27\pm0.43$ &--   &--     &$0.05\pm0.01$ & $0.76\pm0.22$    & yes/no/no      \\
%PNG 081.2+14.9  & A 78     &-5.09 &--   &--            &--            &--              &--            \\
085.4+52.3  & Jacoby 1 & $0.46\pm$-- &327 &-- &$0.01\pm0.01$& $1.00\pm0.29$  & yes/yes/yes    \\
094.0+27.4  & K 1-16   &$0.72\pm0.05$ &47  &--      &$0.38\pm0.01$     &--            & yes/no/no  \\
104.2-29.6  & Jn 1     &$20.\pm$-- & 160.0  &--            &$0.08\pm0.03$ & $1.01\pm0.29$&   no/no/no  \\
105.4-14.0  & Kn 130 $^{2}$  &--    &--   &--            &--            &--              &  no/yes/no    \\
            &          &      &     &              &              &                &                \\
118.8-74.7  & NGC 246  &$29.30\pm03.69$ &112 &200 & $0.02\pm0.01$ &$0.77\pm0.22$ &  yes/yes/yes \\
120.4-01.3  & Ou 2 $^{2}$    &--    &--   &--            &--            &--              & no/no/no              \\
224.3+15.3  & A25 $^{3}$& $0.36\pm$--&166  &--            &$0.03\pm0.02$   &$3.02\pm0.87$  & no/no/no             \\
130.9-10.5  & NGC 650-1&$22.30\pm3.87$&69.2 &370 &$0.14\pm0.04$ &$0.93\pm0.26$ &  yes/yes/yes  \\
131.1+03.9  & BMPJ0739-1418 $^{1}$&$0.33\pm$--&15 &--    &$0.30\pm0.07$    &$3.35\pm0.95$ &  yes/yes/no             \\
147.5+4.7   & FEGU 248-5 $^{1}$ &$1.29\pm$--& 167 &-- &$0.10\pm0.03$  &$2.21\pm0.63$ &   yes/yes/yes   \\
149.7-03.3  & IsWe 1   & $4.55\pm0.43$ &390  &--      &$0.22\pm0.03$    & $0.70\pm0.20$ & no/no/no  \\
152.0+09.5  & StDr 138 $^{3}$ &--    &--   &--            &  $0.33\pm0.03$      &$1.06\pm0.20$  & yes/no/no  \\
164.8+31.1  & Jn Er 1  &$7.72\pm-$&379 &-- &$0.02\pm0.01$& $0.95\pm0.27$ &   no/no/yes \\
205.1+14.2  & A 21     & $42.4\pm3.69$ &307.5  &--            &$0.07\pm0.02$   & $0.45\pm0.13$ & yes/yes/no \\
                &          &      &     &              &              &                &       \\
%PNG 208.5+33.3  & A 30     &-4.91 &--   &--            &--            &--              &--            \\
258.0-15.7  & Wray 17-1& $3.98\pm- $&47 &1000 &-- & $2.30\pm0.66$ &      yes/yes/yes    \\
274.3+9.1   & Lo 4   &$0.32\pm0.04$ & 48 &-- &$0.14\pm0.07$ & $5.61\pm1.65$ & no/no/yes   \\
                &          &      &     &              &              &                &               \\      
$Hybrid$-       &    &           &              &              &                &        &        \\
PG1159-PNe  &           &            &              &              &                &       &      \\
030.6+06.2  & Sh 2-68  &$15.1\pm3.78$  &200   &--    &-- & $1.054\pm-$  & yes/no/no   \\  
036.0+17.6  & A 43     &$0.95\pm0.039$  &40  &--   & $0.17\pm0.13$ & $2.99\pm0.89$  &  yes/yes/no  \\
066.7-28.2  & NGC 7094 & $2.03\pm0.378$ &47 &-- & $0.12\pm0.06$& $2.27\pm0.65$&  yes/yes/yes  \\
%PNG 144.1+06.1  & NGC 1501 &-3.29 &75.0 &3.21$\pm$1.4  &118$\pm$20.0  &60.34$\pm$25.67 &9.04$\pm$1.30    \\
--              & $V* LWLib ^{3}$ &--    &--   &--            &--            &--              &--               \\
                &          &      &     &              &              &                &                 \\           
\hline
\end{tabular}
% \footnote{(a) IR data available at 
(a) IR data available at \url{https://irsa.ipac.caltech.edu/frontpage/}
\end{table*}
%  042.5-14.5  & NGC 6852 & $2.9\pm0.39$  &-- &14 &$0.14\pm0.07$ & $4.65\pm1.33$ &  yes/yes/yes   \\
% 046.8+03.8  & Sh 2-78  & $8.46\pm0.424$ &  -- &300   &  $0.01\pm0.002$ &   $1.054\pm0.20$   & yes/no/no         \\
% 059.7-18.7  & A 72     &$2.20\pm0$ &137 &-- & $0.05\pm0.03$ &$2.56\pm0.73$ &  yes/yes/yes  \\
% 062.4+09.5  & NGC 6765 & $1.73\pm0.405$ &19 &-- &$0.19\pm0.17$ & $3.7\pm1.28$ &  yes/yes/yes  \\
% 070.5+11.0  & Kn 61 $^{2}$   &0.12    &96   &--          & $0.15\pm0.03$   &   $5.41\pm1.51$  & no/no/no  \\
% 080.3-10.4  & MWP1     &$6.27\pm0.434$ &--   &--     &$0.05\pm0.01$ & $0.76\pm0.22$    & yes/no/no            \\
% %PNG 081.2+14.9  & A 78     &-5.09 &--   &--            &--            &--              &--            \\
% 085.4+52.3  & Jacoby 1 & 0.46 &327 &-- &$0.01\pm0.01$& $1.0\pm0.29$  & yes/yes/yes \\

\section{Results}
\label{sec3}
 
 Using the archival photometric fluxes and other nebular parameters collected from the literature, we investigate the IR colours of PG1159- and hybrid PG1159-PNe and derive their physical properties.
 
\subsection{IR Colour-colour Diagrams analysis}

The IR colour-colour diagrams (CCDMs) of PNe are useful tools to study their IR properties. The IR-CCDMs were derived for PG1159- and hybrid PG1159-PNe using their photometric measurements at $2MASS$, $WISE$ and $IRAS$ bands. The IR emission arising from different components of a PN was discussed in MP20 and we use here the same methodology followed by them to obtain the CCDMs, hence they are not discussed here. IR CCDMS of [WR]-,$wels$- and normal-PNe and their mean values of colours, which are used to make a comparison study, are all from MP20. However, we resolve the distribution of [WRL]- and [WRE]-PNe in IR CCDMs to show the evolutionary connection.

\subsubsection{Near-IR colour analysis}

Near-IR radiation of PNe comprises the emission from atoms and free-free and free-bound emission of electrons in the gaseous nebula. The hot dust component of PNe with temperatures $\sim$1000K - 2000K can also radiate significantly in the near-IR waveband \citep{1985MNRAS.213...59W, 1988A&A...202..203L}. We used the near-IR CCDM to analyze the dominant source of radiation at this waveband for PG1159- and hybrid PG1159-PNe and compare them with the other groups of PNe. $2MASS$ photometric measurements were used for all sources in Table 1, as available, to have uniformity in the analysis with other groups of PNe done by MP20. 

The extinction corrected near-IR CCDM of PG1159 and hybrid PG1159-PNe between $2MASS$ bands $[J-H]_{0}$ and $[H-K]_{0}$ is shown in Fig. 1. Also shown in the figure are the regions dominated by emission from stellar component (S-region, this could be a companion to CSPN or field star(s)), nebular hot dust component (D-region) and from nebular atomic, free-free and free-bound emission (N-region). Contribution from both stellar and nebular components are seen in the 'NS' region and from both nebular and dust components are seen in the 'ND' region. More details on different emitting regions in the plot can be seen in Section 3.1 of MP20.  The errors associated with the colours are computed from the errors in the photometric measurements and the errors in the $E(B-V)$ values. Fig. 1 also shows the extinction corrected colours of [WRL]-,[WRE]- $wels$- and normal-PNe in $2MASS$ bands, which are over-plotted for comparison. We find larger errors in the near-IR colours of many of the PG1159 and its hybrid type PNe, when compared to the other groups of PNe, due to the larger errors in their photometric measurements. \\

\begin{figure*}
\includegraphics[width=0.95\textwidth]{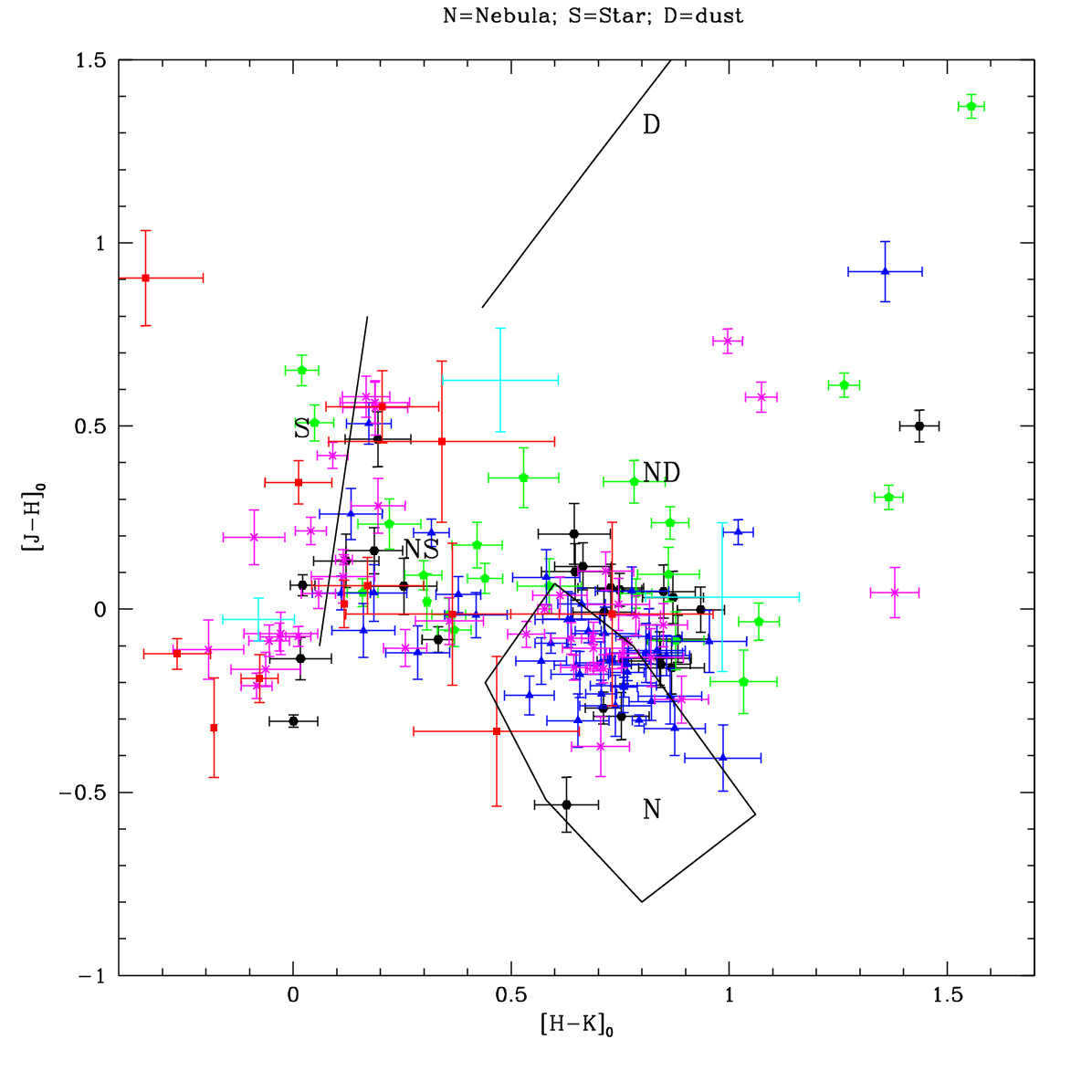}
\caption{Extinction corrected near-IR CCDMs of PG1159- and hybrid PG1159-PNe plotted along with the near-IR colours of [WRE]-,[WRL],$wels$- and normal-PNe computed by MP20. [WRL]-PNe are shown as filled hexagons in black, [WRE]-PNe are as filled pentagons in green, $wels$-PNe are as filled triangles in blue, PG1159-PNe are as filled squares in red, hybrid PG1159-PNe are as filled stars in cyan and normal-PNe are shown as crosses in pink.}
\label{fig:2}        
\end{figure*}

 There are 15 PG1159- and hybrid PG1159-PNe for which the $2MASS$ data and $E(B-V)$ values are available, and we have calculated the near-IR CCDMs for them. Out of these 15 sources, five PG1159-PNe (including one hybrid type) are located in the 'S' region ($\sim 33\%$) and in the 'NS' region there are five PG1159-PNe ($\sim 33\%$). Five PNe are located in the 'ND' region (three PG1159-PNe and two hybrid PG1159-PNe), which represents $\sim$33$\%$ of total. No PG1159 and its hybrid type PNe is seen in the 'N'-region where the He I emission at 1.083$\mu$ dominates \citep{1985MNRAS.213...59W}. As there are only three hybrid PG1159-PNe, we discuss their distribution together with PG1159-PNe. To compare this result, $\sim$ 34$\%$ and $\sim$ 43$\%$ of [WR]-PNe ([WRL] and [WRE]) are positioned in 'S \& NS' and in 'ND' regions, respectively. The distributions of [WRE]- and [WRL]-PNe in the 'S' and 'NS' regions are quite similar ($\sim$ 28\% and $\sim$ 29\% respectively). However, while no [WRL]-PNe is seen in the 'N' region, four [WRE]-PNe are located in this region. In the 'ND' region the distribution of [WRE]-PNe is more closer to the nebular box than the [WRL]-PNe. As the central star becomes hotter, the nebular atomic/ionic emission brightens making more number of [WRE]-PNe in the 'N' region. Otherwise, these two sub-groups of [WR]-PNe are quite similar in the $2MASS$ CCDM. From this analysis, one can find that [WR]-PNe and PG1159-PNe show similar distributions in $2MASS$ CCDMs in the 'S' and 'NS' regions, whereas $wels$-PNe differ from them: they are concentrated in or around the 'N' region. In the 'ND' region [WR]-PNe dominates and PG1159-PNe have their significant presence. It should be noted that A70 and A30 are classified as PNe in their transition phase from [WRE] to PG1159 \citep{weidmann2020catalogue}. But their log($S_{H\beta}$) values derived by us are -5.09 and -4.91 respectively, which are well within the PG1159-PNe range shown in Table 3. These two PNe are located in the 'ND' region on the $2MASS$ CCDM; ($[J-H]_{0}$; $[H-K]_{0}$) values respectively are (0.830; 0.618) for A78 and (0.789; 0.894) for A30. If they are confirmed candidates of PG1159-PNe, then the fraction of PG1159-PNe in the 'ND' region becomes 41$\%$ which is quite similar to the fraction seen for the [WR]-PNe. As noted by MP20, a significant fraction ($\sim$ 31$\%$) of $wels$-PNe is also seen in the 'ND' region. The concentration of more number of PG1159-PNe and [WR]-PNe in the 'ND'-region shows the domination of emission from the hot dust component in these two groups of PNe. This hot dust component is a very small grain population with grain sizes varying from $\sim$ 10 to 50 \AA, which may have formed from the winds of the [WR] CSPN \citep{2001A&A...377.1007G}. They have low heat capacity and hence, they thermally fluctuate in the harsh central star radiation field, transiently attaining very high temperatures (see \cite{10.1093/mnras/stt1319} and \cite{10.1093/mnras/stx1071} for a discussion on this grain population in a dusty PN IRAS18333-2357). PG1159-PNe are faint and have low electron densities due to nebular expansion, which shows that their nebular emissions have declined significantly over time. Hence, they are rarely seen near the 'N' box, whereas 17$\%$ [WR]-PNe are in or around the 'N' box. However, the emission from the hot dust components of PG1159-PNe do not seem to have declined significantly due to the nebular expansion as compared to the nebular line emission. This may be related to the heating of a statistical grain, which is independent of the radial distance from the central star, and the grain number density determines the net emission. Whereas for the gas, the ionizing radiation field is diluted with radial distance from the central star and both the ionization and recombination rates depend on the gas density. It is to be noted that the near-IR colours of normal-PNe are mostly seen near the stellar line, inside or closer to the nebular box and only a few of them ($\sim$16$\%$) are located in the 'ND' region, making a sharp difference with PNe having H-poor CSPNe. \\

\subsubsection{Mid- and Far-IR colour analysis}

 The warm, thermal equilibrium dust component of PN with a typical colour temperature of $\sim$ 150K - 300K was created in the cool and dense atmosphere of the 
 AGB progenitor and is heated by the hard UV radiation of the CSPN \citep{dustmassratio}. This dust component can be traced using the mid-IR bands of $WISE$ and $IRAS$. We use the 
 'N' and 'Q' band fluxes to trace this dust component in PG1159- and hybrid PG1159-PNe and compare them with the other groups of PNe with H-poor central stars to understand their evolution. The 'N' band is also, however, contaminated by the emission from Polycyclic Aromatic Hydrocarbons (PAH) and the atomic 
 emission from forbidden transitions in the nebula (see MP20 for a discussion on this issue). The 22$\mu$m band of $WISE$ and the 25$\mu$m band of $IRAS$ are, however, 
 least contaminated, and they trace the continuum dust better. The mid-IR colours of PG1159 and its hybrid type PNe at $WISE$ bands are shown along with that 
 of [WRL]-, [WRE]-, $wels$- and normal-PNe in Fig. 2 (top panels). Similarly, their CCDMs in the $IRAS$ bands were calculated
 and are shown in Fig. 2, along with the $IRAS$ colours of the other three groups of PNe for a comparison (bottom panels). $IRAS$ fluxes at its 60- 
 and 100-$\mu$m bands are used to trace the cool AGB dust (cooler than 150K) in PNe. Errors in the photometric measurements at their respective wave bands of 
 $WISE$ and $IRAS$ are used to calculate the errors in the colours, which are shown at each point in the figure. As brought out from Fig. 2 the distribution of
 PG1159- and hybrid PG11159-PNe are in the cooler region which is different from that of [WRL]-,[WRE]- and $wels$-PNe in the mid-IR and far-IR CCDMs. Only some normal-PNe are seen in the region which is occupied by PG1159-PNe.
  
 \begin{figure*}
\includegraphics[width=0.95\textwidth]{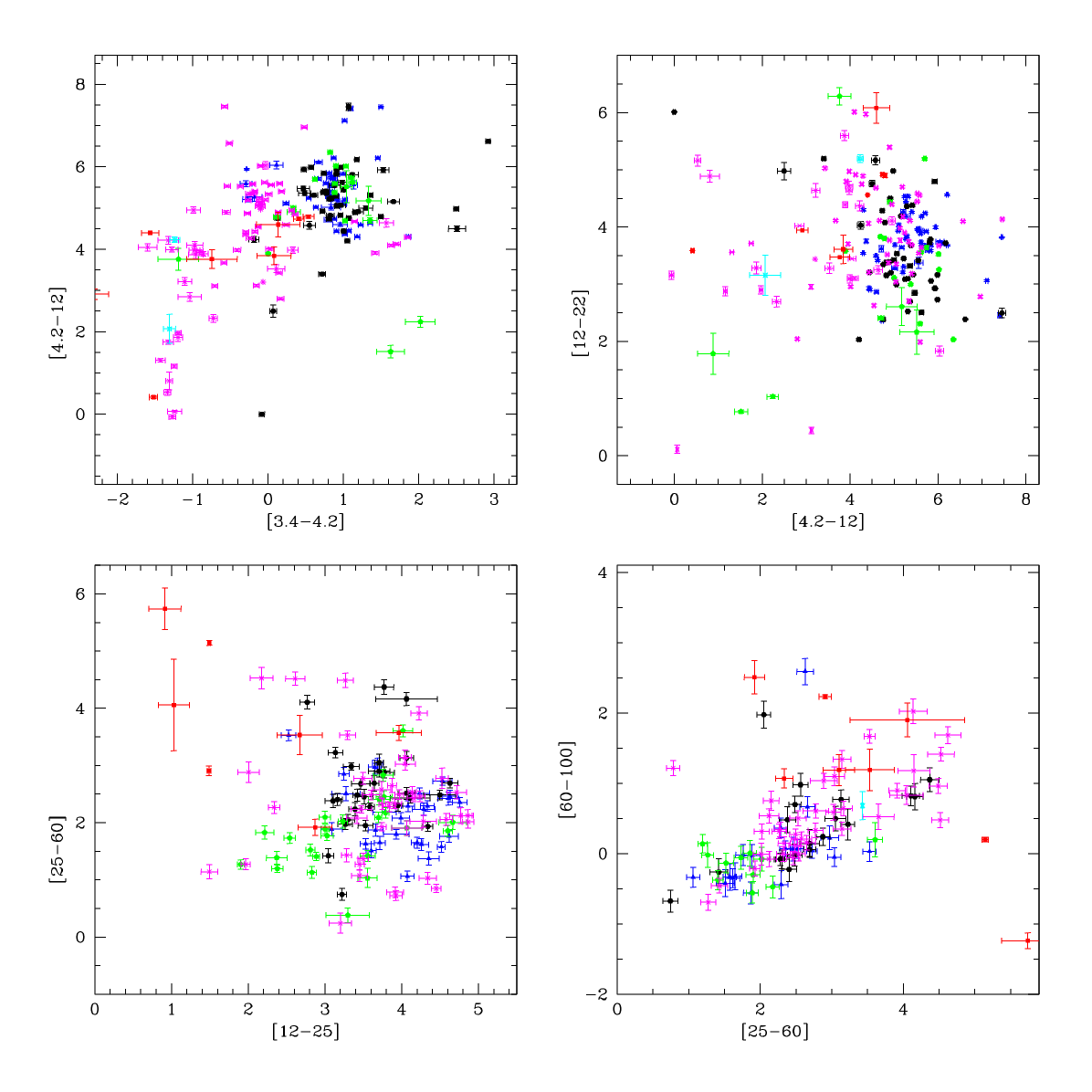}
\caption{Mid-IR CCDMs of PG1159- and hybrid PG1159-PNe derived using $WISE$ (top panels) and $IRAS$ (bottom panels) photometry. They are plotted along with the mid-IR CCDMs of [WRL]-,[WRE]-$wels$- and normal-PNe. Data description as given in Fig. 1.} 
\label{fig:2}       % Give a unique label
\end{figure*}

The mean values of mid-IR colours for [WR]-PNe, $wels$-PNe and normal-PNe are shown with their standard deviations in Table 2 along with 
the values of PG1159-PNe (which includes hybrid PG1159-PNe) derived from this study. As seen in the table, the mean values of $WISE$ colours [3.4-4.2] and [4.2-12] for PG1159-PNe are smaller 
than the values reported for [WR]-, $wels$- and normal-PNe. $WISE$ [12-22] colour for PG1159-PNe is larger than the values obtained for other three
groups; however, the $IRAS$ [12-25] colour shows a significantly smaller value. The number of PG1159-PNe for which the photometric measurements are available 
at $WISE$ 12- and 22-$\mu$m bands are 13, whereas at 12- and 25-$\mu$m bands of $IRAS$, there are only 9 candidates have photometric measurements. Hence, we considered the 
colours derived using the fluxes at $WISE$ 12- and 22-$\mu$m bands for this study. While the $IRAS$ [25-60] colours are quite similar for [WR]-, $wels$- and 
normal-PNe, PG1159-PNe show a larger mean value, which indicates that the AGB dust in PG1159-PNe are relatively cooler than in other groups of PNe. 
Similarly in the far-IR, the mean value of $IRAS$ [60-100] colour for PG1159-PNe from this work is significantly larger than the respective values for 
[WR]- and $wels$-PNe. We find that the mid-IR bands are increasingly brighter as one moves away from the 12$\mu$m band, both below and above this 
wavelength. If we take that the contribution from the atomic/ionic lines relative to the dust continuum emission are similar for all PG1159-PNe, then the above finding 
shows that below 12$\mu$m the continuum emission from the hot dust component dominates for PG1159-PNe and its hybrid type and the cool dust emission dominates above 12$\mu$m. This confirms the finding from $2MASS$ CCDM (Section 3.1.1) that PG1159 and hybrid PG1159-PNe have a significant amount of hot dust component similar to [WR]-PNe. The mid-and far-IR CCDMs in Fig. 2 show 
that PG1159 and hybrid PG1159-PNe are concentrated in a region which is not occupied by the [WR]- and $wels$-PNe and only the faint normal-PNe are seen in this region. This 
clearly shows that PG1159-PNe are more evolved objects than [WR]- and $wels$-PNe, which is also expected from their smaller values of electron densities as
compared to other PNe.

\begin{table}
%\vspace{10cm}
\label{tbl2}
%\begin{tabular}{lrr}
\vspace{0cm}
\caption{{Mean values of colours between different IR bands and their standard deviations for PG 1159-PNe (which also includes hybrid type). They are compared with their respective values of 
[WR]-PNe, $wels$-PNe and normal-PNe. The number of PNe were given between two round brackets.}}
\vspace{0cm}
\setlength{\tabcolsep}{20pt}
\renewcommand{\arraystretch}{1.5}
\begin{tabular}{lrrrr}
\hline
\hline
 Colour   & [WR]-PNe & $wels$-PNe   & PG1159-PNe & Normal-PNe    \\   
\hline
 $WISE$   &          &         &       &      \\
$[3.4-4.2]$   &   0.88$\pm$0.58 (63) & 0.81$\pm$0.29 (47)  & 0.463$\pm$0.52 (11) & 0.77$\pm$0.55 (72) \\
$[4.2-12]$   &  5.17$\pm$1.20 (63)& 5.29$\pm$0.80 (47)      &3.57$\pm$1.1 (10)  & 4.57$\pm$1.50 (71)  \\
$[12-22]$   &  3.37$\pm$0.90(73) & 3.65$\pm$0.55 (55)      &4.30$\pm$0.95 (15)  & 3.65$\pm$1.10 (80)  \\
 $IRAS$   &          &         &                                    \\
$[12-25]$   &  3.45$\pm$0.63 (55)  & 3.95$\pm$0.50 (27)  &  2.06$\pm$0.97 (7)  &  3.73$\pm$0.70 (41)  \\ 
$[25-60]$   &  2.41$\pm$0.74 (78) & 2.38$\pm$0.54 (52)   &  3.45$\pm$1.10 (11) &  2.38$\pm$0.86 (67)  \\
$[60-100]$  &  0.21$\pm$0.50 (29) & 0.17$\pm$0.65 (18)   &  1.12$\pm$1.81 (8)  &  0.68$\pm$0.54 (43)  \\
%            &          &         &            &     \\
%$[65-90]$   &   0.35$\pm$0.37 (50) & 0.38$\pm$0.22 (35) &    & 0.70$\pm$0.50 (53)  \\
%$[90-140]$  &   0.07$\pm$0.70 (45) & 0.34$\pm$0.62 (26)  &   & 0.55$\pm$0.79 (48)  \\
%$[140-160]$  &  -0.11$\pm$1.10 (31)  &-0.18$\pm$0.90 (19)  & -0.46$\pm$1.20 (41) \\
          &         &         &           &    \\
\hline
%\caption{mid- and far-IR emission features of IRAS 18333-2357} 
\end{tabular}
\end{table} 

\subsection{Gas and Dust properties}
  
 We derive the gas and dust properties of PG1159-PNe and hybrid PG1159-PNe as discussed in Section 2. As it can be seen from Table 1. the electron densities are not available for many of the sources in the literature for which we have taken a mean value of 500 cm$^{-3}$, which is a good approximation for PG1159-PNe. We list the values of different parameters, namely, $S_{H\beta}$, $T_{d}$, $M_{d}$, $m_{d}/m_{g}$, $L_{IR}$ and IRE derived for our sample of PG1159-PNe and hybrid PG1159-PNe along with their errors in Table 3.   As there is only one hybrid PG1159-PN for which the gas and dust parameters are available, we discuss its properties together with PG1159-PNe. Table 4. shows the mean values of $T{_d}$, log$(m_{d})$, log$(m_{d}/m_{g})$, log$(L_{IR})$ and log$(IRE)$ and their standard deviations for [WR]-,$wels$- and normal-PNe from MP20 and are compared with PG1159-PNe. The number
 of candidates used to obtain these values are in round brackets. Plots showing variation of different nebular and dust parameters against $S_{H\beta}$ for [WR]-,$wels$- and normal-PNe are from MP20. However, we resolve [WRL]- and [WRE]-PNe in these plots to show the evolutionary connection. We discuss the nebular and dust properties of different groups of PNe below. 
 
 The mean $T_{d}$ value for 11 PG1159-PNe is 75K which is significantly lower than the mean values obtained for other groups of PNe (see Table 4).  We plot the value of $T_{d}$ 
 against $S_{H\beta}$ for PG1159-PNe in Fig. 3 to find its variation as the PN ages. The figure also shows the variation of $T_{d}$ against $S_{H\beta}$ for 
 [WRL]-,[WRE]-, $wels$- and normal-PNe. Fig. 3 shows that all the PG1159-PNe are distributed in the lower end of $S_{H\beta}$ values in the plot, where other PNe with 
 H-poor central stars are not present. However, some of the normal-PNe are seen in this region. As can be seen from the plot, the $T_{d}$ value decreases with 
 $S_{H\beta}$ for PG1159-PNe similar to the trend seen for the other groups of PNe (also see \cite{dusttemp2, 2001A&A...377.1007G}. However, a few PG1159-PNe 
 (and some normal-PNe) show larger temperatures than the typical value of $T_{d}$ expected at this region. The low $S_{H\beta}$ values clearly show that PG1159-PNe 
 are old objects and have expanded to larger sizes as compared to [WR]- and $wels$-PNe. This has resulted in lower gas density and smaller mean $T_{d}$ for them.
 
 \begin{figure}
\includegraphics[width=0.95\textwidth]{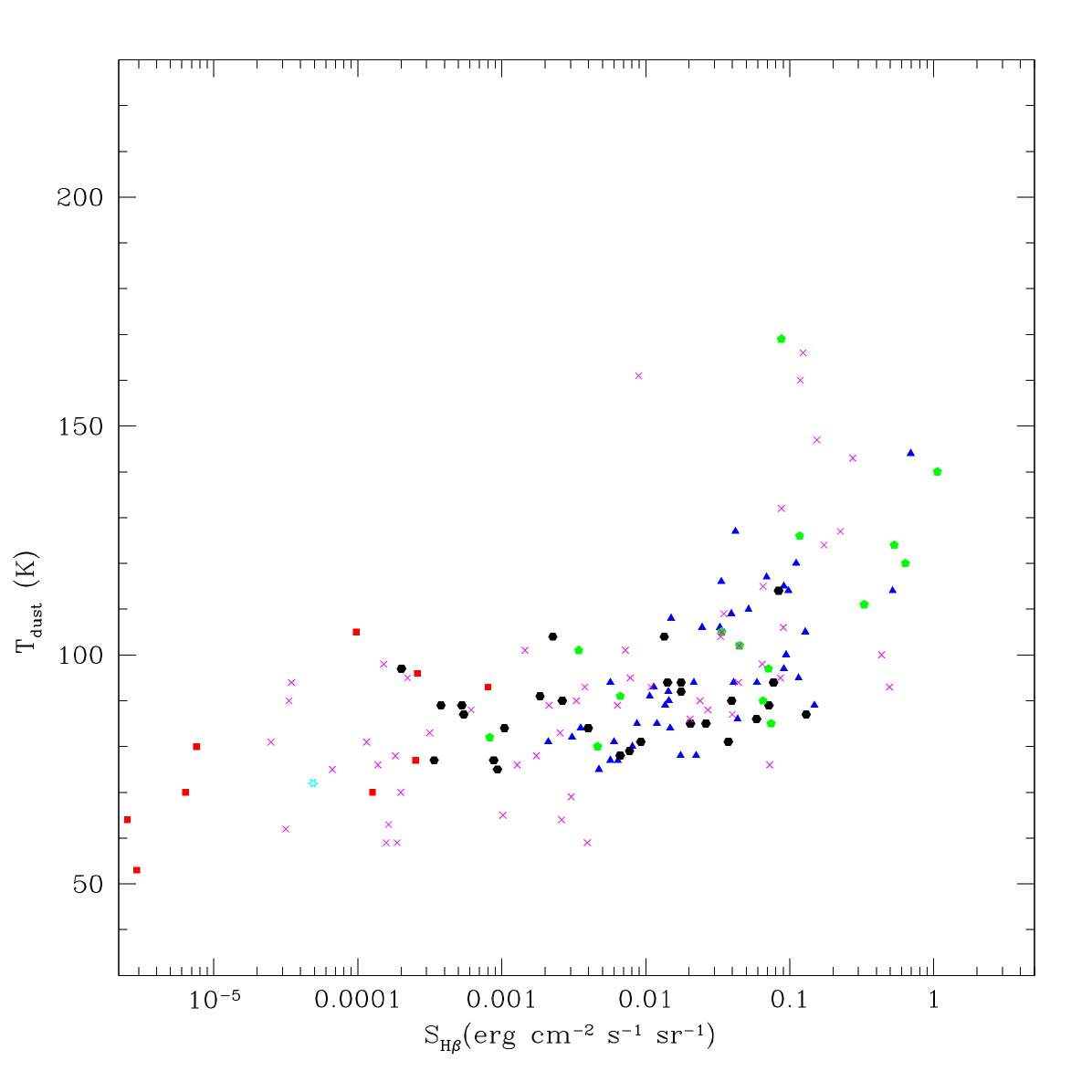}
\caption{Variation of dust colour temperature with $S_{H\beta}$ for PG1159-, and hybrid PG1159-PNe shown with the other groups of PNe derived by MP20. Data description as given in Fig. 1.} 
\label{fig:2}       % Give a unique label
\end{figure}

 The dust mass $m_{d}$ and the dust-to-gas mass ratios $m_{d}/m_{g}$ for 11 PG1159-PNe are plotted against their $S_{H\beta}$ values, and are shown in Fig. 4 (top left and top right panels respectively). 
 The figure also shows the plot for [WRL]-,[WRE]-, $wels$- and normal-PNe. As it can be seen from this plot, while the $m_{d}/m_{g}$ shows no trend with 
 $S_{H\beta}$ for PG1159 and other groups of PNe, the $m_{d}$ plot versus $S_{H\beta}$ is nearly flat down to a value of $S_{H\beta}$ $\sim$ 3 $\times 10^{-3}$ for
 all PNe and it shows a decreasing trend below this value. Most of the PG1159-PNe are located in this region, while the [WRL]-,[WRE]- and $wels$- PNe are predominantly 
 seen in the flat region of the curve. The mean dust mass for PG1159-PNe is significantly lower than the mean values shown by the [WR]-,$wels$- and normal-PNe, as seen in Table 4. However, the mean $m_{d}/m_{g}$ value is marginally larger for PG1159-PNe than the value obtained for the other groups (this is due to a large $m_{d}/m_{g}$ value shown by Jacoby 1, otherwise they are all similar) . The 
  standard deviations associated with the mean values of the parameters and the number of candidates used to derive the mean values for different groups of PNe are shown in the table. 

 IR luminosity ($L_{IR}$) of a PN we refer is the total continuum IR energy emitted from the thermal equilibrium dust (dust emission features are not included). They were estimated for 11 PG1159-PNe as shown in Table 3. The mean value of $L_{IR}$ for PG1159-PNe is significantly lower than the values seen for [WR]-,$wels$- and normal-PNe (see Table 4). A plot of $L_{IR}$ against $S_{H\beta}$ is shown in Fig. 4 (bottom left) for PG1159-PNe along with the other groups of PNe. As it can be seen in the plot, PG1159-PNe follow the same decreasing trend of $L_{IR}$ with $S_{H\beta}$ shown by other PNe groups. However, PG1159-PNe are concentrated at the low $L_{IR}$ regions of the plot as compared to the other PNe with H-poor central stars. The excess amount of $L_{IR}$ in terms of additional IR energy observed from the nebula than what can be accounted by the absorption of Ly$\alpha$ radiation produced in the nebula, which is termed as IR excess (IRE) of a PN. Fig. 4 shows a plot of IRE against $S_{H\beta}$ for PG1159-PNe and also for 
 the other groups of PNe (bottom right). The plot does not show a relation between IRE and $S_{H\beta}$ for all the groups of PNe considered here, implying that IRE does not change as the nebula evolves, as also noted by MP20. $L_{IR}$ and H$\beta$ flux decreases with the age of a PN, keeping the IRE unchanged. As seen from Table 4., PG1159-PNe show a mean IRE value which is lower than that of [WR]-PNe, but still higher than the value shown by the normal PNe. However, the mean value of $L_{IR}$ is significantly lower for PG1159-PNe when compared to the values shown by other groups of PNe.  
 
 \begin{figure*}
\includegraphics[width=0.95\textwidth]{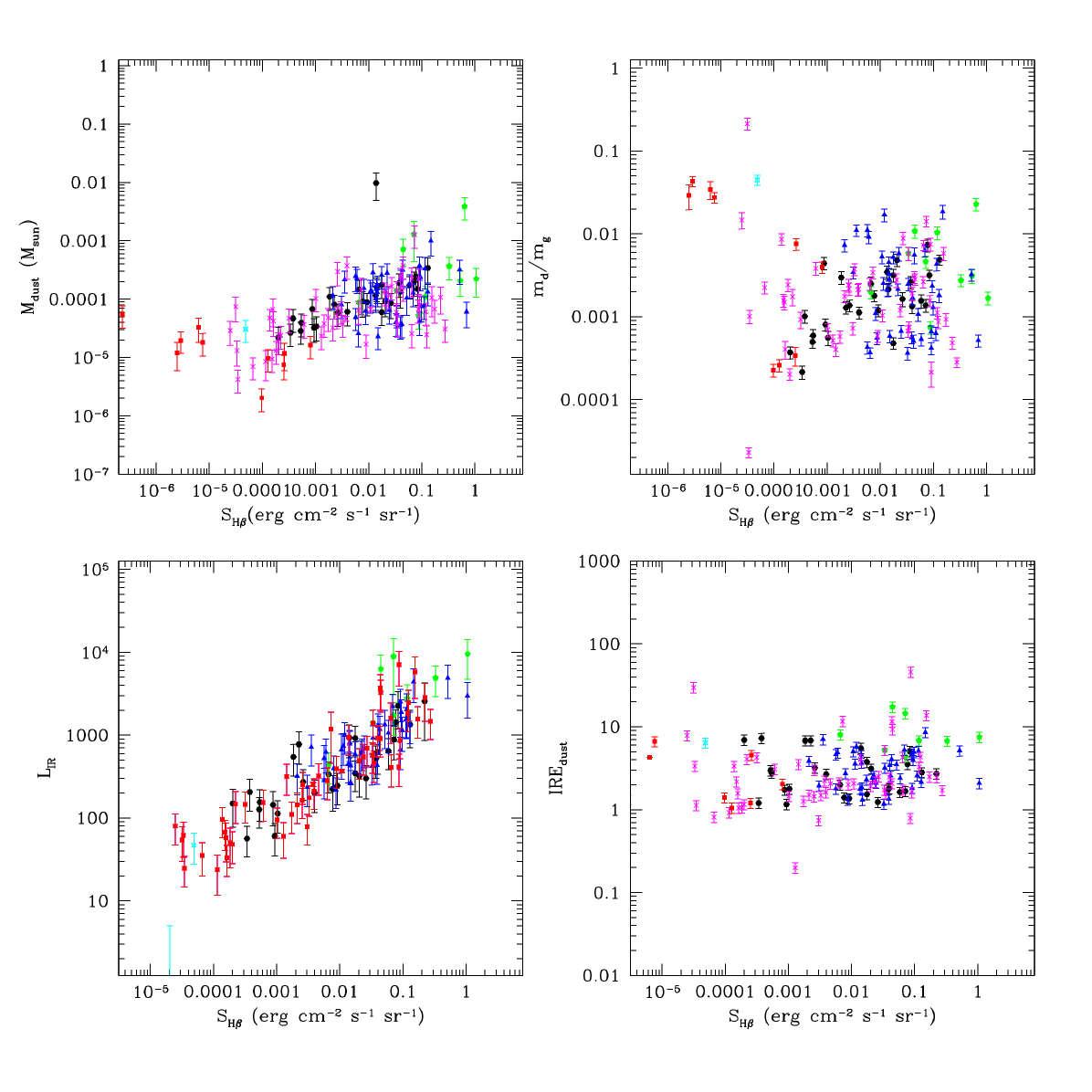}
\caption{Different nebular and dust parameters of PG1159- and hybrid PG1159-PNe derived from this study are plotted against $S_{H\beta}$. Also shown in the plot are [WRL]-,[WRE]-,$wels$- and normal-PNe.  Data description as in Fig. 1.}
\label{fig:2}       % Give a unique label
\end{figure*}

\begin{figure}
\includegraphics[width=0.95\textwidth]{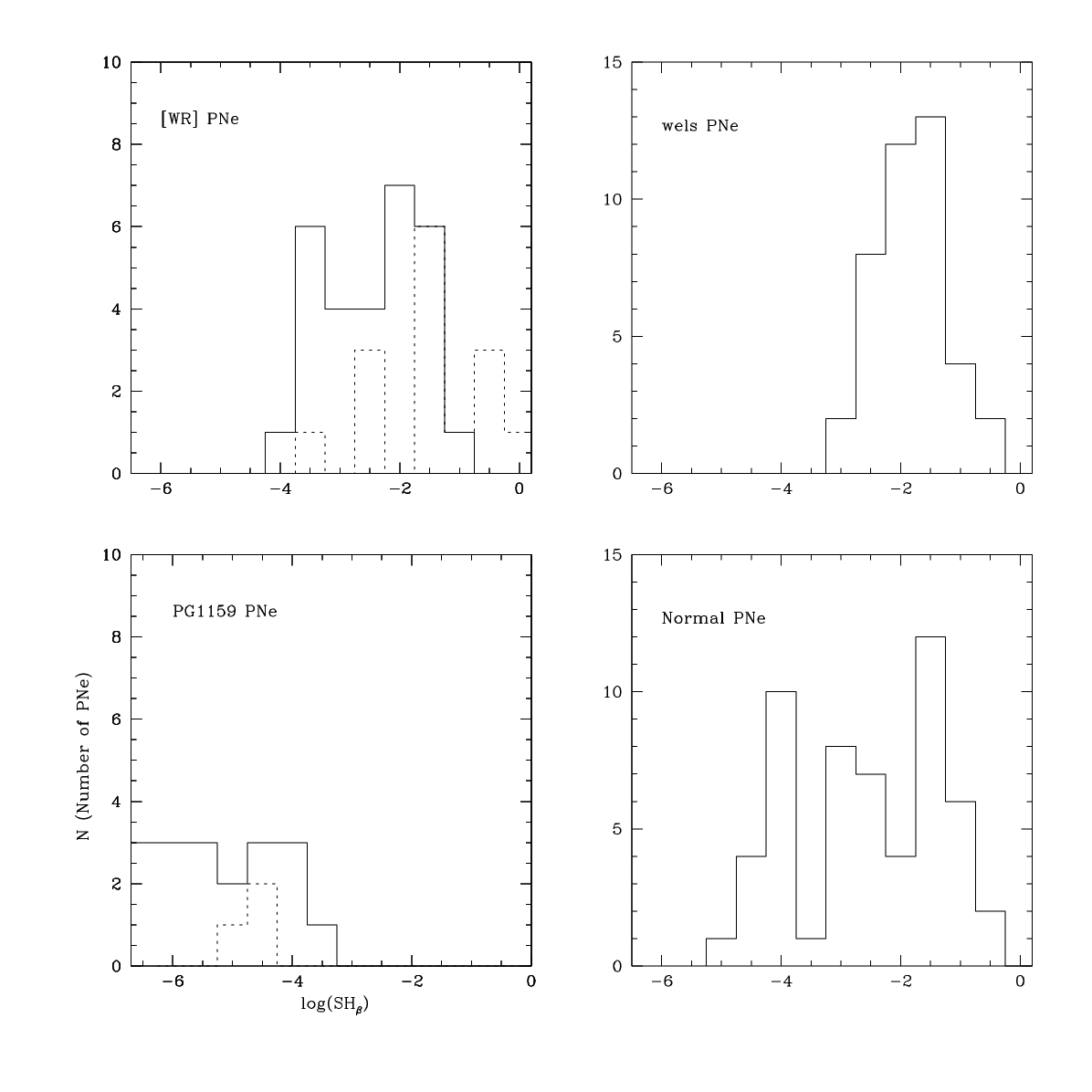}
\caption{Number distribution of PNe against log($S_{H\beta}$) (bin size 0.5).  Top left: [WRL]- (dashed line), [WRE]- (solid line); top right: $wels$-;
bottom left: PG1159- (solid line) and hybrid PG1159- (dashed line) and bottom right: normal-PNe.}
\label{fig:2}       % Give a unique label
\end{figure}

 \begin{table*}
 \centering
%\begin{minipage}{120mm}
\label{tbl1}
\caption{Nebular and dust parameters derived for PG1159 and hybrid PG1159-PNe (see the text for details)}
\vspace{1cm}
\begin{tabular}{@{}lrrrrrrrr}
\hline 
\hline 
PNG &  log($S_{H_{\beta}}$) & $T_{d}$ & $M_{d}$      & $m_{d}/m_{g}$ & $ L_{IR}$        & IRE &    \\
    &                   & (K)     & (10$^{-5}M_\odot$) & (10$^{-3}$) & (L$_{\odot}$)&    &    \\
\hline
                 &      &     &              &              &                &              &   \\
$PG1159-PNe$     &      &     &              &              &                &              &    \\
%PNG 028.0+10.2   &--    &--   &--            &--            &--              &--            &   \\
PNG 042.5-14.5   &-3.10 &93.0 &1.62$\pm$0.66 &3.95$\pm$0.62 &89.33$\pm$36.14 &2.04$\pm$0.29 &    \\
PNG 046.8+03.8   &-5.29 &  -- &--            &  --          &  --            &--            &    \\
PNG 059.7-18.7   &-5.19 &70.0 &3.23$\pm$1.47 &34.40$\pm$8.40 &43.52$\pm$17.48 &4.28$\pm$0.61 &    \\
PNG 062.4+09.5   &-3.58 &96.0 &1.17$\pm$0.58 &7.55$\pm$1.20 &75.49$\pm$39.93 &4.56$\pm$0.64 &    \\
PNG 070.5+11.0   &-6.14   &--   &--            &--            &--              &--            &    \\
PNG 080.3-10.4   &--    &--   &--            &--             &--             &--           &    \\
%PNG 081.2+14.9  & A 78     &-5.09 &--   &--            &--            &--              &--           &    \\
PNG 085.4+52.3   &-6.63 &49.0 &5.45$\pm$2.26 &1811.0$\pm$278&12.18$\pm$5.0   &37.88$\pm$5.36&    \\
PNG 094.0+27.4   &-4.75 &--   &--            &--            &--              &--            &    \\
PNG 104.2-29.6   &-4.37 &--   &--            &--            &--              &--            &    \\
PNG 105.4-14.0   &--    &--   &--            &--            &--              &--            &    \\
                &      &     &              &              &                &              &     \\
PNG 118.8-74.7   &-3.90 &70   &0.96$\pm$0.39 &0.26$\pm$0.05  &12.75$\pm$5.15  &1.05$\pm$0.15 &    \\
PNG 120.4-01.3   &--    &--   &--            &--            &--              &--            &    \\
PNG 224.3+15.3   & -6.16   &--   &--            &--            &--              &--            &    \\
PNG 130.9-10.5   &-3.60 &77.0 &0.76$\pm$0.33 &0.34$\pm$0.09 &16.20$\pm$6.40  &1.20$\pm$0.17 &    \\
PNG 131.1+03.9   & -6.11  &--     &--    &--             &--             &--              &    \\
PNG 147.5+4.7    & -5.60 & 64 & 1.20$\pm$0.61 &29.00$\pm$9.70 & 10.2$\pm$4.13   & 2.3$\pm$0.33  &    \\
PNG 149.7-03.3   &-5.79 &--   &--            &--            &--              &--            &    \\
PNG 152.0+09.5   &--    &--   &--            &--            &--              &--            &    \\
PNG 164.8+31.1   &-5.54 &53.0 &1.96$\pm$0.79 &43.00$\pm$6.10&6.49$\pm$2.61     &1.33$\pm$0.19 &    \\
PNG 205.1+14.2   &-4.61 &--   &--            &--            &--              &--            &    \\
                 &      &     &              &              &                &              &    \\
%PNG 208.5+33.3  & A 30     &-4.91 &--   &--            &--            &--              &--            &   \\
PNG 258.0-15.7   &-4.01 &105.0&0.20$\pm$0.09 &0.22$\pm$0.04 &20.58$\pm$8.35  &1.4$\pm$0.20  &    \\
PNG 274.3+9.1    &-5.12 &80.0 &1.82$\pm$0.76 &27.00$\pm$3.90 &47.26$\pm$19.7  &6.67$\pm$0.94 &    \\
                 &      &     &              &              &                &              &     \\      
$Hybrid$ $PG1159-PNe$ &    &            &              &              &                &              &     \\
PNG 030.6+06.2   &-4.69 &--   &--            &--            &--              &--            &     \\  
PNG 036.0+17.6   &-4.49 &--   &--            &--            &--              &--            &     \\
PNG 066.7-28.2   &-4.31 &72.0 &3.06$\pm$1.24 &44.70$\pm$6.30  &46.87$\pm$18.98 &6.41$\pm$0.91 &     \\
%PNG 144.1+06.1  & NGC 1501 &-3.29 &75.0 &3.21$\pm$1.4  &118$\pm$20.0  &60.34$\pm$25.67 &9.04$\pm$1.30 &     \\
He 1429-1209     &--   &--   &--            &--            &--              &--                   \\
                 &      &     &              &              &                &              &     \\           
\hline
\end{tabular}
\end{table*}

\subsection{$S_{H\beta}$ analysis of PNe}

There are 15 [WRL]-, 29 [WRE]- , 42 $wels$-, and 55 normal-PNe for which the values of $S_{H\beta}$ were given by MP20 and we have derived the $S_{H\beta}$ values for 18 PG1159-PNe and 3 hybrid PG1159-PNe in this work. From this data, we plot the number distribution of PNe against $S_{H\beta}$ values for [WRL]-,[WRE]-, $wels$-, PG1159-, hybrid PG1159- and normal-PNe. We
considered a bin size of 0.5 for log($S_{H\beta}$), and the histograms for all groups of PNe considered in this study are shown in Fig. 5.  We perform a
statistical analysis on the distributions of PNe against $S_{H\beta}$, which can provide useful information on the origin and the evolutionary connection 
between the different groups of PNe considered in this study.

Fig. 5 shows that the number distribution of PG1159-PNe and hybrid PG1159-PNe are distinctly different from the other groups: they are seen at very low $S_{H\beta}$ values while 
the other groups of PNe are seen at the higher values of $S_{H\beta}$. In addition, PG1159-PNe show good overlap with [WRE]-PNe in this plot. Similarly [WRL]-PNe are seen at lower values of $S_{H\beta}$ than [WRE]-PNe and they also show a good overlap. This indicates an evolutionary sequence of [WRL]-PNe $\rightarrow$ [WRE]-PNe $\rightarrow$ PG1159-PNe. Hybrid PG1159-PNe are located within the $S_{H\beta}$ range shown by other PG1159-PNe, indicating that hybrid PG1159-PNe do not form an evolutionary link between PG1159-PNe and other PNe with H-poor central stars. However, it is not apparent from the figure to say if the number distributions for [WR]-,$wels$- 
and normal-PNe are similar or different. To trace this, we invoke the Kolmogorov-Smirnov (KS) test statistics, which has been popularly used to check if the two given data sets 
 (which are independent of their number distributions) are similar or differ significantly. The KS test is particularly useful for comparing cumulative distribution 
functions (CDFs) of two data sets. \\
%\textbf{The empirical CDF(ECDFs) can be calculated at a point x of the dataset using the relation:\\
%\[ECDF = \frac{Number\ of\ observations (\leq x)}{Total\ number\ of\ observations}\]}

The statistics is parameterized by two values: the 'D' value of KS statistics shows the maximum absolute difference between the CDFs of two data sets, and the 'P' value of the statistics 
indicates the probability of the null hypothesis that the distributions of the two data sets are the same \citep{10.1093/mnras/202.3.615, TEEGAVARAPU20191}. A low 'D' value shows that 
the difference between the two data sets is very low (D $\ge$ 0.5 infers significantly different data sets) and a high 'P'  value shows that the hypothesis can not be 
rejected. We perform the KS test between the number distribution of [WR]-,$wels$- and normal-PNe with $S_{H\beta}$ values to see if these data were derived from a 
similar sample. We find D = 0.14 and P=0.99 between the distributions of [WR]- and normal-PNe if one takes the entire data set or we take only the first peak (below
$S_{H\beta}$ = 0.001, which represents the younger group of PNe).  KS test between [WRL]- and normal-PNe gives the values of D= 0.2 and P=1.0. Between $wels$- and normal-PNe distributions, D = 0.28 and P = 0.63 if one takes the full data set 
and D = 0.14 and P = 0.99 if we take the first peak. Our analysis shows that, particularly at higher values of $S_{H\beta}$ (i.e., at the early stages of PNe), the 
number distribution functions of [WRL]-PNe and $wels$-PNe against $S_{H\beta}$ are very similar to that of normal-PNe and hence they were derived from a similar sample. 
%\textbf{The KS test between the number distribution of [WRE]- and PG1159-PNe shows D = 0.18 and P = 0.95. This implies that [WRL]- and PG1159-PNe are evolutionarily connected; required ?}.

% if one takes the entire data set or we take only the first peak (below
% $S_{H\beta}$ = 0.001, which represents the younger group of PNe). 
%Between $wels$- and PG1159-PNe, D = 0.36 and P = 0.25, showing significant differences in the distributions. Our analysis shows that, particularly at higher values of $S_{H\beta}$ (i.e., at the early stages of PNe), the 
%number distribution functions of [WR]-PNe against $S_{H\beta}$ are very similar to that of PG1159-PNe and hence they were derived from a similar sample.}
 
\begin{figure}
\includegraphics[width=0.95\textwidth]{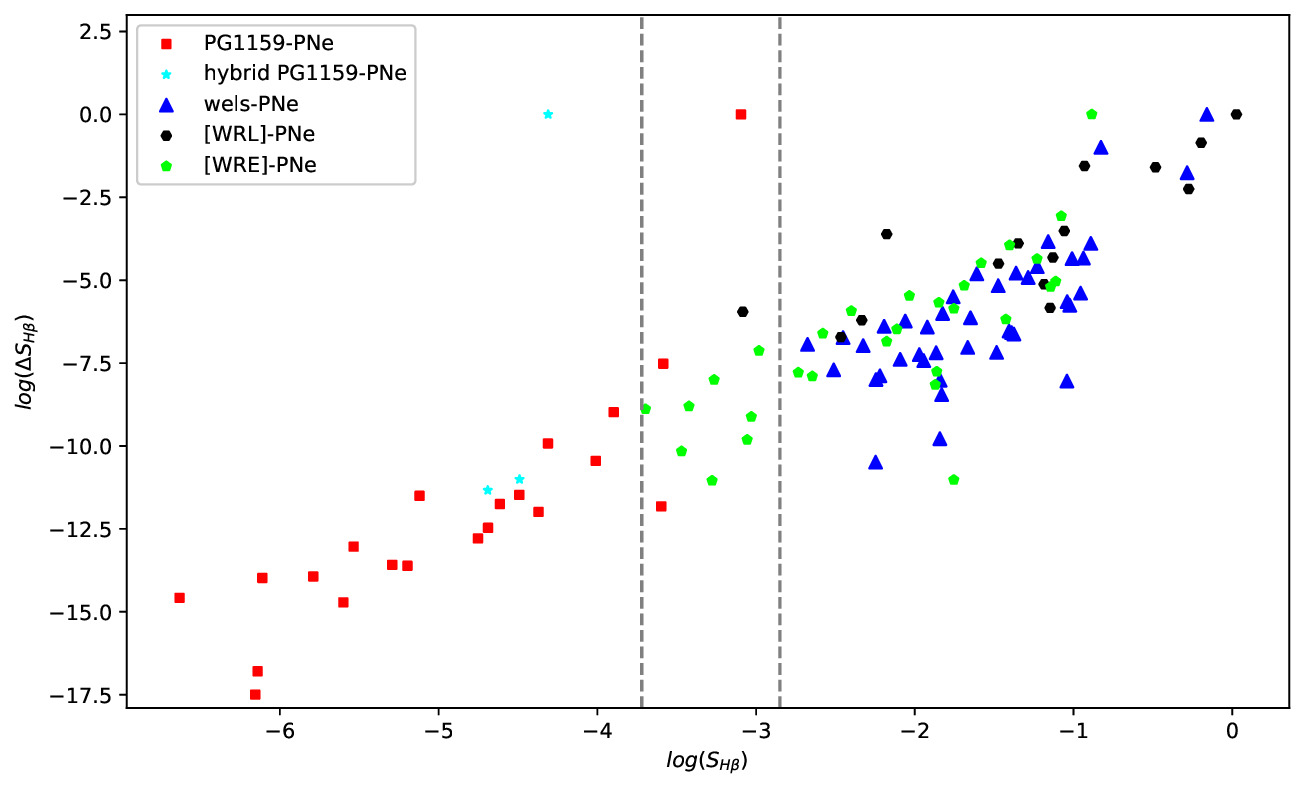}
\caption{A plot of log$(\Delta S_{H\beta}$) versus log $S_{H\beta}$ for different group of PNe with H-poor CSPNe. See the text for details.} 
\label{fig:2}       % Give a unique label
\end{figure}

Fig. 6 shows the difference between subsequent $S_{H\beta}$ values for PNe plotted against their $S_{H\beta}$ values, both in logarithmic scale, for [WRL]-, [WRE]-, 
$wels$-, PG1159- and hybrid PG1159-PNe. This plot clearly shows that there is a good overlap between [WRE]- and PG1159-PNe in $S_{H\beta}$ indicating an evolutionary connection. A weak overlap is seen between [WRL]- and PG1159-PNe as [WRL]-PNe evolve to [WRE]-PNe.  However, the $wels$-PNe have a significant 
gap in $S_{H\beta}$ with the PG1159-PNe in this plot. In the logarithmic scale the gap in log($S_{H\beta}$) is 0.42, which is 3 times the mean of the differences in successive values of log($S_{H\beta}$) for $wels$-PNe (0.13) and 2.5 times that for PG1159-PNe (0.17). However, it should be noted that the number of PG1159-PNe 
known till date are not many and there is also a possibility that this gap in log($S_{H\beta}$) could be due to the lack of observational points in this range.

\begin{table}[htp!]
\centering
\begin{tabular}{l|c|c|c|c}
    \hline
    \hline
         Parameter & [WR]-PNe & \textit{wels}-PNe & PG-1159 & Normal PNe \\
          (mean value)    &(No. of PNe) & (No. of PNe) &  (No. of PNe) & (No. of PNe)  \\
    \hline
                    &                  &                  &                    &                \\
         $T_{d}$(K) & $96 \pm25 (78) $ & $95 \pm 14 (52)$ &   $75 \pm 17 (11)$ & $93 \pm 31 (67)$ \\
         log[$m_{d_{\odot}}$] & $-3.83 \pm 0.51 (70)$ & $-3.86 \pm 0.35 (49)$ & $-4.71 \pm 0.24 (11)$ & $-4.33 \pm 0.56 (61)$ \\
         log[$m_d/m_g$] & $-2.38 \pm 0.63$ (52)& $ -2.37 \pm 0.50$ (43)& $-2.05 \pm0.51$ (11) & $-2.51 \pm 0.58$ (48) \\
         log[$L_{IR}(L\odot)$] & $2.91 \pm 0.55$ (67) & $2.90 \pm 0.32$ (49) & $ 1.53\pm 0.23$ (11) & $2.76 \pm 0.50$ (61) \\
         log[$IRE$] & $0.64 \pm 0.60$ (54) & $0.47 \pm 0.22$ (42)& $ 0.50 \pm 0.22$ (11) & $0.43 \pm 0.40$ (52) \\
       \hline  
\end{tabular}
    \caption{ Mean values of nebular and dust parameters and their standard deviations for PG1159-PNe obtained from this study are shown along with their respective 
    values for [WR]-,$wels$- and normal-PNe. See the text for more details.}.
    \label{mean_values}
\end{table}

 \section{Discussion}
 \label{sec4}
   From the results we obtained above using IR data and $S_{H\beta}$ analysis, we make a few suggestions on the evolution of [WR]-,$wels$-, PG1159-PNe and discuss them in the following sections.

  \subsection{Dust-to-gas mass ratios of PNe}
   
   Does the grain population in PN undergo substantial evolution during its evolution and does the $m_{d}/m_{g}$ ratio of PN decrease before the envelope is 
   dispersed into the ISM were key questions which were debated for a long time. \cite{1981ApJ...248..189N, article} have discussed this elaborately in their analytical 
   study made on 243 PNe using $IRAS$ data and have suggested that the mean grain size and $m_{d}/m_{g}$ show significant changes during the evolution of PNe. \cite{iraspaper} 
   also arrived at the same conclusion using $IRAS$ photometry of 46 PNe. \cite{stasinskaed} used extensive photo-ionization models for 500 PNe, constrained by $IRAS$ fluxes, and found no evidence for the decrease of $m_{d}/m_{g}$ with the age of the nebula. They also found that the decrease in the mean grain size found by \cite{article} is an artefact associated with their analysis and that the timescale of grain destruction is larger than the lifetime of a PN. Fig. 4 (top left panel) shows that the dust mass is not correlated 
   with $S_{H\beta}$ down to a value of $S_{H\beta}$ $\sim$ 3 $\times$ 10$^{-3}$ for all PNe and the graph shows a decreasing trend below this value. Most of the PG1159-PNe and some 
   normal-PNe are located in this region. This is, however, not a real picture as found in the models of \cite{stasinskaed}. They brought out that the atomic emission lines 
   of [OIII] at 25.9$\mu$m and 51.8$\mu$m can contaminate significantly the fluxes at 25- and 60$\mu$m bands of $IRAS$. Hence, the dust colour temperature derived from the 
   fluxes at these two bands could be overestimated, leading to a smaller dust mass. As also seen from Fig. 3, larger dust colour temperatures are seen at low $S_{H\beta}$ values, supporting this argument. We find a wide range in the values of $m_{d}/m_{g}$ for all PNe up to a factor of $\sim$50; most PNe have values in Fig.4 (top right panel) from 0.0002 to 0.01. However,
   partly this could be caused by two reasons: 1) we have assumed an emissivity exponent of grains $\alpha$ = -1 uniformly for all PNe. While this value suits for 
   carbonaceous grains, for silicates a value of $\alpha$ = -2 will be more appropriate. This overestimates the dust colour temperature and in turn underestimates 
   $m_{d}/m_{g}$. Our analysis shows that taking carbonaceous grains in the place of silicates will lead to a maximum difference of 40$\%$ in the estimated value of $m_{d}/m_{g}$. Also, one can note that most of [WR]- and $wels$-PNe and about 50$\%$ of normal-PNe have carbonaceous grains, as discussed by MP20. 2)  some PNe are radiation-bounded (about 20$\%$;\cite{stasinskaed}) and have significant amount of neutral mass around ionized region, and the gas mass estimated from H$\beta$ flux refers to only the ionized gas component. This leads to an underestimation of the total nebular mass (ionized and neutral gas), in turn overestimating $m_{d}/m_{g}$. We arrive at the same conclusion of \cite{stasinskaed} that PNe are formed
   with a range of $m_{d}/m_{g}$ and in addition we also do see that this is irrespective of if their CSPNe are H-rich or H-poor.
   
  \subsection{Origin of PNe with H-rich and H-poor central stars}

  It was believed that the H-poor atmosphere of [WR] star was formed due to thermal pulse occurring at the final-AGB phase, or late (while the star evolves in the 
  constant luminosity track after ejection of PN) or very late (when the CSPNe evolve through the WD cooling track) AGB phases (born-again PNe; \cite{10.1007/978-94-011-2088-3_271, ref2}. PNe, which have gone through late- or very late-thermal pulse phases show resolved envelopes in their optical images 
  which were ejected at two different epochs of the progenitor star's evolution. For example, the born-again PNe A30 and A78 \citep{1994ApJ...435..722B, 2015ASPC..493..141T}. However, most of the [WR]- or $wels$-PNe do not show such morphologies in their resolved optical images. Our finding that the young PNe with [WR] and $wels$ central 
  stars have the number distribution functions against $S_{H\beta}$ which are very similar to that of the normal-PNe implies that PNe of these three groups originated in a 
  quite similar way. This indicates that [WR]- and $wels$- PNe were formed directly from AGB evolution as how the normal-PNe were born without facing late or very late 
  thermal pulses. There are, however, a very few exceptions: there are five born-again PNe known till date, and among them, a few show [WR] type central star like the 
  CSPNe of A58 and IRAS 15154-5258 \citep{2002Ap&SS.279..171Z} which have seen the late thermal pulse. The overlap of $S_{H\beta}$  values in Fig. 6 between [WRE]- and PG1159-PNe shows a 
  continuity in $S_{H\beta}$ and hence an evolutionary connection between them and [WRE] and [WRL] are evolutionarily connected. Such a continuity is not seen between $wels$- and PG1159-PNe. This may indicate that 
  $wels$-PNe and PG1159-PNe may not form an evolutionary sequence, though this may require more observations for confirmation. While none of the [WR] stars show hydrogen in their spectra, hybrid-PG1159 stars do show hydrogen in absorption.
  %examples for hybrid-PG1159 CSPNe are the central stars of PNe Abell 43 and NGC 7094 \citep{ref2}.
  This may indicate that some PG1159 stars can also be formed through other rout(s) which are different from [WR] phenomena. 

 \section{Conclusions}
 \label{sec5}
 
  We studied the properties of PNe around 26 PG1159 stars known till date, which includes four hybrid PG1159 type, and compared them with the properties of [WR]-, $wels$- and normal-PNe reported earlier to 
  understand their evolutionary status. We used archival IR photometry of PNe in addition to their H$\beta$ fluxes, sizes and distances obtained from the literature,
  as available, to make this study. We make the following conclusions:
  \begin{enumerate}
  \item The near-IR CCDMs of PG1159 and hybrid PG1159-PNe show a tendency of their distribution near the stellar line (66\%). Next to this they tend to populate in the 'ND' region which is dominated by the [WR]-PNe. This may indicate the presence of significant amount of hot dust (statistically fluctuating small grains) within the nebula. The AGB dust traced by the mid-IR bands are relatively cooler for PG1159-PNe when compared to [WR]-, $wels$- and normal-PNe.
  \item PG1159 and its hybrid type PNe are found at the lower end of $S_{H\beta}$ where [WR]-PNe and $wels$-PNe are not seen and only some normal-PNe are present. 
  smaller IR luminosities and similar IRE when compared to other groups. While their dust-to-gas mass ratio is somewhat larger than that of the other groups of PNe, their mean dust mass is smaller. The dust-to gas mass ratio is not correlated with $S_{H\beta}$ for all groups of PNe.
  \item There is a good overlap between [WRE]- and PG1159-PNe in their $S_{H\beta}$ values which indicates an evolutionary connection between them. However, the $wels$-PNe show significant gap in $S_{H\beta}$ with PG1159-PNe. The number distributions of [WRL]- and $wels$-PNe against $S_{H\beta}$ are very similar with that of PNe with H-rich central stars. This may indicate that [WR]-,$wels$- and normal PNe have evolved in a similar way, i.e. direct evolution from the AGB.
  \end{enumerate}
 
 \section{Acknowledgements}
 \label{sec6}

This research has made use of the SIMBAD database, operated at CDS, Strasbourg, France. This publication makes use of data products from the Two Micron 
All Sky Survey, which is a joint project of the University of Massachusetts and the Infrared Processing and Analysis Centre, funded by the National 
Aeronautics and Space Administration and the National Science Foundation. This work is based in part on observations made with $WISE$, obtained from the 
NASA/IPAC Infrared Science Archive, operated by the Jet Propulsion Laboratory, California Institute of Technology, under contract with the National 
Aeronautics and Space Administration. The authors thank the anonymous referee for his/her useful comments which have improved the manuscript. The research work was supported by the DST-SERB grant of Govt. of India (Grant No: CRG/2020/000755) received by 
the author C. Muthumariappan.

%% Bibliography
%% Author year style
\bibliography{muthu}

\begin{thebibliography}{45}
\providecommand{\natexlab}[1]{#1}
\providecommand{\url}[1]{\texttt{#1}}
\expandafter\ifx\csname urlstyle\endcsname\relax
  \providecommand{\doi}[1]{doi: #1}\else
  \providecommand{\doi}{doi: \begingroup \urlstyle{rm}\Url}\fi

\bibitem[{Acker} and {Neiner}(2003)]{2003A&A...403..659A}
A.~{Acker} and C.~{Neiner}.
\newblock {Quantitative classification of WR nuclei of planetary nebulae}.
\newblock \emph{aap}, 403:\penalty0 659--673, May 2003.
\newblock \doi{10.1051/0004-6361:20030391}.

\bibitem[{Awad} and {Ali}(2023)]{2023RAA....23i5021A}
Z.~{Awad} and A.~{Ali}.
\newblock {Physical and Kinematical Characteristics of Wolf-Rayet Central Stars and their Host Planetary Nebulae}.
\newblock \emph{Research in Astronomy and Astrophysics}, 23\penalty0 (9):\penalty0 095021, September 2023.
\newblock \doi{10.1088/1674-4527/acd993}.

\bibitem[{Bond} et~al.(2023){Bond}, {Werner}, {Jacoby}, and {Zeimann}]{2023MNRAS.521..668B}
Howard~E. {Bond}, Klaus {Werner}, George~H. {Jacoby}, and Gregory~R. {Zeimann}.
\newblock {Spectroscopic survey of faint planetary-nebula nuclei - I. Six new 'O VI' central stars}.
\newblock \emph{mnras}, 521\penalty0 (1):\penalty0 668--676, May 2023.
\newblock \doi{10.1093/mnras/stad524}.

\bibitem[{Borkowski} et~al.(1994){Borkowski}, {Harrington}, {Blair}, and {Bregman}]{1994ApJ...435..722B}
Kazimierz~J. {Borkowski}, J.~Patrick {Harrington}, William~P. {Blair}, and Jesse~D. {Bregman}.
\newblock {The Dust in the Hydrogen-poor Ejecta of Abell 30}.
\newblock \emph{The Astrophysical Journal}, 435:\penalty0 722, November 1994.
\newblock \doi{10.1086/174849}.

\bibitem[{Catelan} et~al.(1996){Catelan}, {de Freitas Pacheco}, and {Horvath}]{thermalpulse}
M.~{Catelan}, J.~A. {de Freitas Pacheco}, and J.~E. {Horvath}.
\newblock {The Helium-Core Mass at the Helium Flash in Low-Mass Red Giant Stars: Observations and Theory}.
\newblock \emph{The Astrophysical Journal}, 461:\penalty0 231, April 1996.
\newblock \doi{10.1086/177051}.

\bibitem[{Crowther}(2008)]{2008ASPC..391...83C}
P.~A. {Crowther}.
\newblock {Physical and Wind Properties of [WC] Stars}.
\newblock In A.~{Werner} and T.~{Rauch}, editors, \emph{Hydrogen-Deficient Stars}, volume 391 of \emph{Astronomical Society of the Pacific Conference Series}, page~83, July 2008.
\newblock \doi{10.48550/arXiv.0710.5774}.

\bibitem[{Cutri}(2012)]{allwise}
R.~M.~{et al.} {Cutri}.
\newblock {VizieR Online Data Catalog: WISE All-Sky Data Release (Cutri+ 2012)}.
\newblock \emph{VizieR Online Data Catalog}, art. II/311, April 2012.

\bibitem[{De Marco}(2008)]{2008ASPC..391..209D}
O.~{De Marco}.
\newblock {[WC] and PG 1159 Central Stars of Planetary Nebulae: The Need for an Alternative to the Born-Again Scenario}.
\newblock In A.~{Werner} and T.~{Rauch}, editors, \emph{Hydrogen-Deficient Stars}, volume 391 of \emph{Astronomical Society of the Pacific Conference Series}, page 209, July 2008.
\newblock \doi{10.48550/arXiv.0711.2461}.

\bibitem[{Frew} et~al.(2016){Frew}, {Parker}, and {Boji{\v{c}}i{\'c}}]{frew2016halpha}
David~J. {Frew}, Q.~A. {Parker}, and I.~S. {Boji{\v{c}}i{\'c}}.
\newblock {The H{\ensuremath{\alpha}} surface brightness-radius relation: a robust statistical distance indicator for planetary nebulae}.
\newblock \emph{Monthly Notices of the Royal Astronomical Society}, 455\penalty0 (2):\penalty0 1459--1488, January 2016.
\newblock \doi{10.1093/mnras/stv1516}.

\bibitem[Girard et~al.(2007)Girard, Koppen, and Acker]{girard:hal-00119849}
Pascal Girard, J.~Koppen, and A.~Acker.
\newblock {Chemical compositions and plasma parameters of planetary nebulae with Wolf-Rayet and wels type central stars}.
\newblock \emph{{Astronomy \& Astrophysics}}, 463:\penalty0 265--274, 2007.
\newblock URL \url{https://hal.science/hal-00119849}.
\newblock 17 pages, 28 figures,.

\bibitem[{G{\'o}rny} and {Tylenda}(2000)]{2000A&A...362.1008G}
S.~K. {G{\'o}rny} and R.~{Tylenda}.
\newblock {Evolutionary status of hydrogen-deficient central stars of planetary nebulae}.
\newblock \emph{Astronomy \& Astrophysics}, 362:\penalty0 1008--1019, October 2000.

\bibitem[{G{\'o}rny} et~al.(2001){G{\'o}rny}, {Stasi{\'n}ska}, {Szczerba}, and {Tylenda}]{2001A&A...377.1007G}
S.~K. {G{\'o}rny}, G.~{Stasi{\'n}ska}, R.~{Szczerba}, and R.~{Tylenda}.
\newblock {Infrared properties of planetary nebulae with [WR] central stars}.
\newblock \emph{Astronomy \& Astrophysics}, 377:\penalty0 1007--1015, October 2001.
\newblock \doi{10.1051/0004-6361:20011050}.

\bibitem[{Green} et~al.(1986){Green}, {Schmidt}, and {Liebert}]{greensurvey}
R.~F. {Green}, M.~{Schmidt}, and J.~{Liebert}.
\newblock {The Palomar-Green Catalog of Ultraviolet-Excess Stellar Objects}.
\newblock \emph{ApJS}, 61:\penalty0 305, June 1986.
\newblock \doi{10.1086/191115}.

\bibitem[Iben(1993)]{10.1007/978-94-011-2088-3_271}
Icko Iben.
\newblock The evolution of planetary nebulae, their precursors and their progeny --- a commentary.
\newblock In Ronald Weinberger and Agnes Acker, editors, \emph{Planetary Nebulae}, pages 587--596, Dordrecht, 1993. Springer Netherlands.
\newblock ISBN 978-94-011-2088-3.

\bibitem[{Leene} and {Pottasch}(1988)]{1988A&A...202..203L}
A.~{Leene} and S.~R. {Pottasch}.
\newblock {IRAS pointed observations of planetary nebulae.}
\newblock \emph{aap}, 202:\penalty0 203--214, August 1988.

\bibitem[Lenzuni et~al.(1989)Lenzuni, Natta, and Panagia]{article}
Paolo Lenzuni, Antonella Natta, and N.~Panagia.
\newblock Properties and evolution of dust grains in planetary nebulae.
\newblock \emph{The Astrophysical Journal}, 345:\penalty0 306--326, 09 1989.
\newblock \doi{10.1086/167906}.

\bibitem[L\"{o}bling et~al.(2019)L\"{o}bling, Rauch, Bertolami, Todt, Friederich, Ziegler, Werner, and Kruk]{ref2}
L~L\"{o}bling, T~Rauch, M~M~Miller Bertolami, H~Todt, F~Friederich, M~Ziegler, K~Werner, and J~W Kruk.
\newblock Spectral analysis of the hybrid {PG}{\hspace{0.167em}}1159-type central stars of the planetary nebulae abell{\hspace{0.167em}}43{\hspace{0.167em}}and {NGC}{\hspace{0.167em}}7094.
\newblock \emph{Monthly Notices of the Royal Astronomical Society}, 489\penalty0 (1):\penalty0 1054--1071, July 2019.
\newblock \doi{10.1093/mnras/stz1994}.
\newblock URL \url{https://doi.org/10.1093/mnras/stz1994}.

\bibitem[Löbling et~al.(2019)Löbling, Rauch, Miller Bertolami, Todt, Friederich, Ziegler, Werner, and Kruk]{10.1093/mnras/stz1994}
L~Löbling, T~Rauch, M~M Miller Bertolami, H~Todt, F~Friederich, M~Ziegler, K~Werner, and J~W Kruk.
\newblock {Spectral analysis of the hybrid PG 1159-type central stars of the planetary nebulae Abell 43 and NGC 7094}.
\newblock \emph{Monthly Notices of the Royal Astronomical Society}, 489\penalty0 (1):\penalty0 1054--1071, 07 2019.
\newblock ISSN 0035-8711.
\newblock \doi{10.1093/mnras/stz1994}.
\newblock URL \url{https://doi.org/10.1093/mnras/stz1994}.

\bibitem[{Mendez}(1991)]{1991IAUS..145..375M}
R.~H. {Mendez}.
\newblock {Photospheric Abundances in Central Stars of Planetary Nebulae, and Evolutionary Implications}.
\newblock In Georges {Michaud} and A.~V. {Tutukov}, editors, \emph{Evolution of Stars: the Photospheric Abundance Connection}, volume 145, page 375, January 1991.

\bibitem[Muthumariappan(2017)]{10.1093/mnras/stx1071}
C.~Muthumariappan.
\newblock {Three-dimensional Monte Carlo dust radiative transfer study of the H-poor planetary nebula IRAS 18333–2357 located in M22}.
\newblock \emph{Monthly Notices of the Royal Astronomical Society}, 470\penalty0 (1):\penalty0 626--638, 06 2017.
\newblock ISSN 0035-8711.
\newblock \doi{10.1093/mnras/stx1071}.

\bibitem[{Muthumariappan} and {Parthasarathy}(2020)]{cmuthu}
C.~{Muthumariappan} and M.~{Parthasarathy}.
\newblock {Infrared properties of planetary nebulae with [WR] and wels central stars}.
\newblock \emph{Monthly Notices of the Royal Astronomical Society}, 493\penalty0 (1):\penalty0 730--746, March 2020.
\newblock \doi{10.1093/mnras/staa217}.

\bibitem[Muthumariappan et~al.(2006)Muthumariappan, Kwok, and Volk]{Muthumariappan_2006}
C.~Muthumariappan, Sun Kwok, and Kevin Volk.
\newblock Subarcsecond mid-infrared imaging of dust in the bipolar nebula hen 3-401*.
\newblock \emph{The Astrophysical Journal}, 640\penalty0 (1):\penalty0 353, mar 2006.
\newblock \doi{10.1086/500041}.
\newblock URL \url{https://dx.doi.org/10.1086/500041}.

\bibitem[Muthumariappan et~al.(2013)Muthumariappan, Parthasarathy, and Ita]{10.1093/mnras/stt1319}
C.~Muthumariappan, M.~Parthasarathy, and Y.~Ita.
\newblock {Radiative transfer modelling of dust in IRAS 18333−2357: the only planetary nebula in the metal-poor globular cluster M22}.
\newblock \emph{Monthly Notices of the Royal Astronomical Society}, 435\penalty0 (1):\penalty0 606--622, 08 2013.
\newblock ISSN 0035-8711.
\newblock \doi{10.1093/mnras/stt1319}.

\bibitem[{Natta} and {Panagia}(1981)]{1981ApJ...248..189N}
A.~{Natta} and N.~{Panagia}.
\newblock {Dust in planetary nebulae.}
\newblock \emph{The Astrophysical Journal}, 248:\penalty0 189--194, August 1981.
\newblock \doi{10.1086/159142}.

\bibitem[{Neugebauer} et~al.(1984){Neugebauer}, {Habing}, {van Duinen}, {Aumann}, {Baud}, {Beichman}, {Beintema}, {Boggess}, {Clegg}, {de Jong}, {Emerson}, {Gautier}, {Gillett}, {Harris}, {Hauser}, {Houck}, {Jennings}, {Low}, {Marsden}, {Miley}, {Olnon}, {Pottasch}, {Raimond}, {Rowan-Robinson}, {Soifer}, {Walker}, {Wesselius}, and {Young}]{iraspaper}
G.~{Neugebauer}, H.~J. {Habing}, R.~{van Duinen}, H.~H. {Aumann}, B.~{Baud}, C.~A. {Beichman}, D.~A. {Beintema}, N.~{Boggess}, P.~E. {Clegg}, T.~{de Jong}, J.~P. {Emerson}, T.~N. {Gautier}, F.~C. {Gillett}, S.~{Harris}, M.~G. {Hauser}, J.~R. {Houck}, R.~E. {Jennings}, F.~J. {Low}, P.~L. {Marsden}, G.~{Miley}, F.~M. {Olnon}, S.~R. {Pottasch}, E.~{Raimond}, M.~{Rowan-Robinson}, B.~T. {Soifer}, R.~G. {Walker}, P.~R. {Wesselius}, and E.~{Young}.
\newblock {The Infrared Astronomical Satellite (IRAS) mission.}
\newblock \emph{The Astrophysical Journal, Letters}, 278:\penalty0 L1--L6, March 1984.
\newblock \doi{10.1086/184209}.

\bibitem[Peacock(1983)]{10.1093/mnras/202.3.615}
J.~A. Peacock.
\newblock {Two-dimensional goodness-of-fit testing in astronomy}.
\newblock \emph{Monthly Notices of the Royal Astronomical Society}, 202\penalty0 (3):\penalty0 615--627, 03 1983.
\newblock ISSN 0035-8711.
\newblock \doi{10.1093/mnras/202.3.615}.

\bibitem[{Pottasch} et~al.(1988){Pottasch}, {Bignell}, {Olling}, and {Zijlstra}]{dusttemp2}
S.~R. {Pottasch}, C.~{Bignell}, R.~{Olling}, and A.~A. {Zijlstra}.
\newblock {Planetary nebulae near the galactic center. I. Method of discovery and preliminary results.}
\newblock \emph{Astronomy \& Astrophysics}, 205:\penalty0 248--256, October 1988.

\bibitem[{Schoenberner} and {Napiwotzki}(1990)]{ref10}
D.~{Schoenberner} and R.~{Napiwotzki}.
\newblock {Spectroscopic investigation of old planetaries. I. Detection of two new ``PG 1159'' central stars.}
\newblock \emph{Astronomy \& Astrophysics}, 231:\penalty0 L33--L35, May 1990.

\bibitem[{Skrutskie} et~al.(2006){Skrutskie}, {Cutri}, {Stiening}, {Weinberg}, {Schneider}, {Carpenter}, {Beichman}, {Capps}, {Chester}, {Elias}, {Huchra}, {Liebert}, {Lonsdale}, {Monet}, {Price}, {Seitzer}, {Jarrett}, {Kirkpatrick}, {Gizis}, {Howard}, {Evans}, {Fowler}, {Fullmer}, {Hurt}, {Light}, {Kopan}, {Marsh}, {McCallon}, {Tam}, {Van Dyk}, and {Wheelock}]{2mass}
M.~F. {Skrutskie}, R.~M. {Cutri}, R.~{Stiening}, M.~D. {Weinberg}, S.~{Schneider}, J.~M. {Carpenter}, C.~{Beichman}, R.~{Capps}, T.~{Chester}, J.~{Elias}, J.~{Huchra}, J.~{Liebert}, C.~{Lonsdale}, D.~G. {Monet}, S.~{Price}, P.~{Seitzer}, T.~{Jarrett}, J.~D. {Kirkpatrick}, J.~E. {Gizis}, E.~{Howard}, T.~{Evans}, J.~{Fowler}, L.~{Fullmer}, R.~{Hurt}, R.~{Light}, E.~L. {Kopan}, K.~A. {Marsh}, H.~L. {McCallon}, R.~{Tam}, S.~{Van Dyk}, and S.~{Wheelock}.
\newblock {The Two Micron All Sky Survey (2MASS)}.
\newblock \emph{aj}, 131\penalty0 (2):\penalty0 1163--1183, February 2006.
\newblock \doi{10.1086/498708}.

\bibitem[{Stasi{\'n}ska} and {Szczerba}(1999{\natexlab{a}})]{stasinskaed}
G.~{Stasi{\'n}ska} and R.~{Szczerba}.
\newblock {VizieR Online Data Catalog: The dust content of planetary nebulae (Stasi{\'n}ska+, 1999)}.
\newblock \emph{VizieR Online Data Catalog}, art. J/A+A/352/297, November 1999{\natexlab{a}}.

\bibitem[{Stasi{\'n}ska} and {Szczerba}(1999{\natexlab{b}})]{dustmassratio}
Gra{\.z}yna {Stasi{\'n}ska} and Ryszard {Szczerba}.
\newblock {The dust content of planetary nebulae: a reappraisal}.
\newblock \emph{Astronomy \& Astrophysics}, 352:\penalty0 297--307, December 1999{\natexlab{b}}.
\newblock \doi{10.48550/arXiv.astro-ph/9911006}.

\bibitem[Teegavarapu(2019)]{TEEGAVARAPU20191}
Ramesh~S.V. Teegavarapu.
\newblock Chapter 1 - methods for analysis of trends and changes in hydroclimatological time-series.
\newblock In Ramesh Teegavarapu, editor, \emph{Trends and Changes in Hydroclimatic Variables}, pages 1--89. Elsevier, 2019.
\newblock ISBN 978-0-12-810985-4.
\newblock \doi{https://doi.org/10.1016/B978-0-12-810985-4.00001-3}.
\newblock URL \url{https://www.sciencedirect.com/science/article/pii/B9780128109854000013}.

\bibitem[{Todt} et~al.(2015){Todt}, {Guerrero}, {Fang}, {Toala}, {Arthur}, {Blair}, {Chu}, {Gruendl}, {Hamann}, {Marquez-Lugo}, {Oskinova}, {Ruiz}, {Steffen}, and {Schoenberner}]{2015ASPC..493..141T}
H.~{Todt}, M.~A. {Guerrero}, X.~{Fang}, J.~A. {Toala}, J.~S. {Arthur}, W.~P. {Blair}, Y.~H. {Chu}, R.~A. {Gruendl}, W.~R. {Hamann}, R.~A. {Marquez-Lugo}, L.~{Oskinova}, N.~{Ruiz}, M.~{Steffen}, and D.~{Schoenberner}.
\newblock {The Born-again Planetary Nebulae Abell 30 and Abell 78}.
\newblock In P.~{Dufour}, P.~{Bergeron}, and G.~{Fontaine}, editors, \emph{19th European Workshop on White Dwarfs}, volume 493 of \emph{Astronomical Society of the Pacific Conference Series}, page 141, June 2015.

\bibitem[{Todt} et~al.(2006){Todt}, {Gr{\"a}fener}, and {Hamann}]{2006IAUS..234..127T}
Helge {Todt}, G{\"o}tz {Gr{\"a}fener}, and Wolf-Rainer {Hamann}.
\newblock {Revised element abundances for WC-type central stars}.
\newblock In Michael~J. {Barlow} and Roberto~H. {M{\'e}ndez}, editors, \emph{Planetary Nebulae in our Galaxy and Beyond}, volume 234, pages 127--130, January 2006.
\newblock \doi{10.1017/S1743921306002869}.

\bibitem[{Weidmann} and {Gamen}(2011)]{2011A&A...526A...6W}
W.~A. {Weidmann} and R.~{Gamen}.
\newblock {Central stars of planetary nebulae: New spectral classifications and catalogue}.
\newblock \emph{aap}, 526:\penalty0 A6, February 2011.
\newblock \doi{10.1051/0004-6361/200913984}.

\bibitem[{Weidmann} et~al.(2023){Weidmann}, {Werner}, {Ahumada}, {Pignata}, and {Firpo}]{2023A&A...676A...1W}
W.~A. {Weidmann}, K.~{Werner}, J.~A. {Ahumada}, R.~A. {Pignata}, and V.~{Firpo}.
\newblock {Revealing the ionising star of evolved planetary nebulae}.
\newblock \emph{aap}, 676:\penalty0 A1, August 2023.
\newblock \doi{10.1051/0004-6361/202346401}.

\bibitem[Weidmann et~al.(2020)Weidmann, Mari, Schmidt, Gaspar, Bertolami, Oio, Guti{\'e}rrez-Soto, Volpe, Gamen, and Mast]{weidmann2020catalogue}
Walter~Alfredo Weidmann, MB~Mari, Eduardo~Osvaldo Schmidt, Gaia Gaspar, MM~Miller Bertolami, Gabriel~Andr{\'e}s Oio, LA~Guti{\'e}rrez-Soto, Maria~Gabriela Volpe, R~Gamen, and Damian Mast.
\newblock Catalogue of the central stars of planetary nebulae-expanded edition.
\newblock \emph{Astronomy \& Astrophysics}, 640:\penalty0 A10, 2020.

\bibitem[Werner(1993)]{Werner1993}
K.~Werner.
\newblock \emph{PG 1159 Stars and Related Objects}, pages 67--75.
\newblock Springer Netherlands, Dordrecht, 1993.
\newblock ISBN 978-94-011-2020-3.
\newblock \doi{10.1007/978-94-011-2020-3_10}.
\newblock URL \url{https://doi.org/10.1007/978-94-011-2020-3_10}.

\bibitem[{Werner} et~al.(2004){Werner}, {Rauch}, {Napiwotzki}, {Christlieb}, {Reimers}, and {Karl}]{2004A&A...424..657W}
K.~{Werner}, T.~{Rauch}, R.~{Napiwotzki}, N.~{Christlieb}, D.~{Reimers}, and C.~A. {Karl}.
\newblock {Identification of a DO white dwarf and a PG 1159 star in the ESO SN Ia progenitor survey (SPY)}.
\newblock \emph{aap}, 424:\penalty0 657--663, September 2004.
\newblock \doi{10.1051/0004-6361:20041157}.

\bibitem[Werner and Herwig(2006)]{Werner_2006}
Klaus Werner and Falk Herwig.
\newblock The elemental abundances in bare planetary nebula central stars and the shell burning in agb stars.
\newblock \emph{Publications of the Astronomical Society of the Pacific}, 118\penalty0 (840):\penalty0 183, feb 2006.
\newblock \doi{10.1086/500443}.
\newblock URL \url{https://dx.doi.org/10.1086/500443}.

\bibitem[{Whitelock}(1985)]{1985MNRAS.213...59W}
P.~A. {Whitelock}.
\newblock {JHK photometry of planetary nebulae.}
\newblock \emph{Monthly Notices of the Royal Astronomical Society}, 213:\penalty0 59--69, March 1985.
\newblock \doi{10.1093/mnras/213.1.59}.

\bibitem[{Winget} et~al.(1985){Winget}, {Kepler}, {Robinson}, {Nather}, and {Odonoghue}]{pg1159winget}
D.~E. {Winget}, S.~O. {Kepler}, E.~L. {Robinson}, R.~E. {Nather}, and D.~{Odonoghue}.
\newblock {A measurement of secular evolution in the pre-white dwarf star PG 1159-035.}
\newblock \emph{The Astrophysical Journal}, 292:\penalty0 606--613, May 1985.
\newblock \doi{10.1086/163193}.

\bibitem[{Woudt} et~al.(2012){Woudt}, {Warner}, and {Zietsman}]{2012MNRAS.426.2137W}
Patrick~A. {Woudt}, Brian {Warner}, and Ewald {Zietsman}.
\newblock {SDSS J0349-0059 is a GW Virginis star}.
\newblock \emph{Monthly Notices of the Royal Astronomical Society}, 426\penalty0 (3):\penalty0 2137--2141, November 2012.
\newblock \doi{10.1111/j.1365-2966.2012.21899.x}.

\bibitem[{Wright} et~al.(2010){Wright}, {Eisenhardt}, {Mainzer}, {Ressler}, {Cutri}, {Jarrett}, {Kirkpatrick}, {Padgett}, {McMillan}, {Skrutskie}, {Stanford}, {Cohen}, {Walker}, {Mather}, {Leisawitz}, {Gautier}, {McLean}, {Benford}, {Lonsdale}, {Blain}, {Mendez}, {Irace}, {Duval}, {Liu}, {Royer}, {Heinrichsen}, {Howard}, {Shannon}, {Kendall}, {Walsh}, {Larsen}, {Cardon}, {Schick}, {Schwalm}, {Abid}, {Fabinsky}, {Naes}, and {Tsai}]{wisepaper}
Edward~L. {Wright}, Peter R.~M. {Eisenhardt}, Amy~K. {Mainzer}, Michael~E. {Ressler}, Roc~M. {Cutri}, Thomas {Jarrett}, J.~Davy {Kirkpatrick}, Deborah {Padgett}, Robert~S. {McMillan}, Michael {Skrutskie}, S.~A. {Stanford}, Martin {Cohen}, Russell~G. {Walker}, John~C. {Mather}, David {Leisawitz}, III {Gautier}, Thomas~N., Ian {McLean}, Dominic {Benford}, Carol~J. {Lonsdale}, Andrew {Blain}, Bryan {Mendez}, William~R. {Irace}, Valerie {Duval}, Fengchuan {Liu}, Don {Royer}, Ingolf {Heinrichsen}, Joan {Howard}, Mark {Shannon}, Martha {Kendall}, Amy~L. {Walsh}, Mark {Larsen}, Joel~G. {Cardon}, Scott {Schick}, Mark {Schwalm}, Mohamed {Abid}, Beth {Fabinsky}, Larry {Naes}, and Chao-Wei {Tsai}.
\newblock {The Wide-field Infrared Survey Explorer (WISE): Mission Description and Initial On-orbit Performance}.
\newblock \emph{The Astronomical Journal}, 140\penalty0 (6):\penalty0 1868--1881, December 2010.
\newblock \doi{10.1088/0004-6256/140/6/1868}.

\bibitem[{Zijlstra}(2002)]{2002Ap&SS.279..171Z}
Albert~A. {Zijlstra}.
\newblock {Hydrogen-poor planetary nebulae}.
\newblock \emph{Astrophysics and Space Science}, 279:\penalty0 171--182, January 2002.
\newblock \doi{10.48550/arXiv.astro-ph/0105448}.

\end{thebibliography}

\end{document}